\documentclass{aastex62}
\usepackage{amsmath}

\received{June 12, 2018}
\revised{October 7, 2018}
\accepted{October 15, 2018}
\submitjournal{ApJ}

\shorttitle{the GD1 stream}
\shortauthors{Li, G.-W. et al.}

\begin{document}

\title{GD-1: the relic of an old metal-poor globular cluster}

\email{lgw@bao.ac.cn}

\author{Guang-Wei Li}
\affiliation{Key Laboratory for Optical Astronomy, National Astronomical Observatories, Chinese Academy of Sciences, Beijing 100101, China}

\author{Brian Yanny}
\affil{Fermi National Accelerator Laboratory, Batavia, IL 60510, USA}

\author{Yue Wu}
\affiliation{National Astronomical Observatories, Chinese Academy of Sciences, Beijing 100101, China}



\begin{abstract}
Combining data from Gaia DR2, SDSS DR14 and LAMOST DR6, we update the fit to model of the properties of the stellar stream GD-1 and find that it has an age of $\sim 13$ Gyr, [Fe/H] of $-2.2\pm 0.12$, and a distance from the sun of $\sim 8$ kpc. We tabulate 6D phase-space fiducial points along the GD-1 stream orbit over a 90$^\circ$ arc. The fitted orbit shows that the stream has an eccentricity $e \sim 0.3$, perigalacticon of 14.2 kpc, apogalacticon of 27.0 kpc and inclination $i \sim 40^{\circ}$. There is evidence along the arc for 4 candidate stellar overdensities, one candidate gap, two candidate stellar underdensities and is cut off at $\phi_1 \sim 2^{\circ}$ (in the stream-aligned $(\phi_1,\phi_2)$ coordinate system of \citet{kop10}). The spur originating at $\phi_1 \sim -40^{\circ}$ implies stars were pulled away from the stream trace by an encounter (potentially a dark matter subhalo).  The narrowest place ( FWHM $\sim 44.6$ pc) of the GD-1 trace is at $(\phi_1, \phi_2^c) \sim (-14^{\circ}, 0.15^{\circ})$, which is $\sim (178.18^{\circ}, 52.19^{\circ})$ in (R.~A., Decl.), where the progenitor is possibly located. We also find six BHB and 10 BS spectroscopic stars in the GD-1 stream.

\end{abstract}



\section{Introduction} \label{sec:intro}
The stellar stream GD-1 was discovered by \citet{gri06}. It was traced over $63^{\circ}$ on the sky
(and now is known to extend over $\sim 90^{\circ}$) but is only $\sim 0.5^{\circ}$ wide.  Using velocity 
and metallicity measures of stream members from the 
Sloan Extension for Galactic Understanding and Exploration (SEGUE) survey \citep{yan09}, \citet{wil09} 
fit a retrograde orbit of perigalacticon 14.4 kpc, apogalacticon 
28.7 kpc, inclination $i \sim 35^{\circ}$ and [Fe/H] $= -2.1 \pm 0.1$. Later, \citet{kop10} fit 
a 6D phase space map of the stream, which strongly constrains the circular velocity at the Sun's radius 
and the shape of the Galactic potential (also see \citet{bow15,bov16}). \citet{kop10} also noted 
stellar density fluctuations along the stream, and conjectured that the clumps and holes may be related 
to either the history of the disruption process or interaction of the stream stars with dark 
matter subhalos around 
the Milk Way.  \citet{car13} and \citet{car16} interpreted these gaps as massive dark matter 
subhalo encounters with this cold stellar stream. Recently, \citet{pri18} find two big gaps and 
one spur along the GD-1 trace. The gap at $\phi_1 \sim -45^{\circ}$ is also reported by 
\citet{boe18}. We also notice the work of \cite{hua18}, but our work is different from it in both methods and contents.
\par
In this paper, we refine GD-1 stream parameters and discuss all the topics mentioned above by combining 
data from Gaia DR2, SDSS DR14 and LAMOST DR6. This paper is organized as following: 
In Section \ref{sec:data}, we will present the color-magnitude diagram(CMD), metallicity, 
6D phase-space along the trace; Then the orbit fitting is practiced in Section \ref{sec:orbit}; 
The stellar density fluctuation along the stream trace and blue horizontal branch(BHB) stars, 
and blue stragglers(BS) are discussed in Section \ref{sec:discuss}; 
Finally, conclusions are given in Section \ref{sec:con}.

\section{Data} \label{sec:data}
The data from Gaia DR2\citep{gai18,lin18} and SDSS DR9 are crossed 
match using the database \textsl{gaiadr2.sdssdr9\_best\_neighbour} in TopCat \citep{tay05}. 
Parameters ($T_{\mathrm{eff}}$, $\log g$, [Fe/H] and radial velocity (RV)) of the spectra from LAMOST DR6\footnote{\url{http://dr6.lamost.org}}
\citep{2012RAA....12.1197C,2012RAA....12..723Z} are calculated by LASP (LAMOST Stellar Parameter Pipeline)\citep{wu11a, wu14}, and the main algorithm used in LASP is ULySS\footnote{\url{http://ulyss.univ-lyon1.fr}}\citep{kol08,kol09} with ELODIE interpolator\citep{wu11b}. We also use ULYSS with ELODIE interpolator to recalculate parameters of the spectra from SDSS DR14, so all spectral parameters used in this paper are all from the same pipeline. In the following, magnitudes with subscript 0 indicate they have been 
corrected by the extinction given by \citet{sch11}, which is 0.86 times those given by 
\citet{sch98}, and we also denote $gr_0 \equiv g_0 - r_0$. We adopt in this paper the stream-centered 
$(\phi_1, \phi_2)$ coordinate system and conversion equations given by \citet{kop10}.
There are several papers in the literature which document a systematic underestimate of Gaia DR2 parallaxes, with varying shifts from -0.029 to -0.07 to -0.08 mas \citep{lin18,zetal18,st18}. In this paper we adjust Gaia DR2 parallaxes by adding by 0.029 mas.

\subsection{Astrometric and Photometric Data}
\begin{figure}[ht!]
\plotone{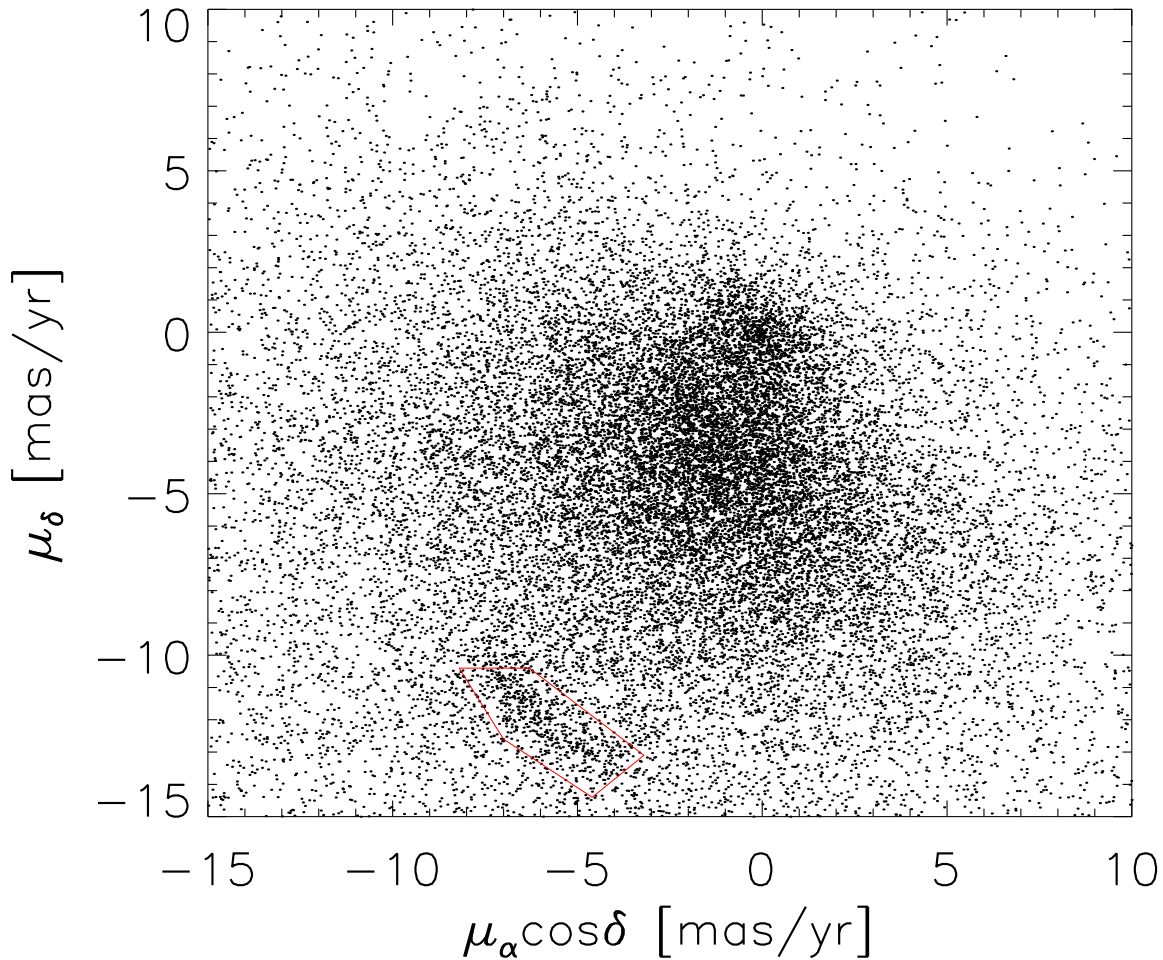}
\caption{The proper motions of stars within $ -60^{\circ} < \phi_1 < -20^{\circ}$ and 
$-0.2^{\circ} < \phi_2 < 0.2^{\circ}$. The red polygon is the region where we select GD-1 
member candidates.\label{photo_pm}}
\end{figure}

\begin{figure}

\plotone{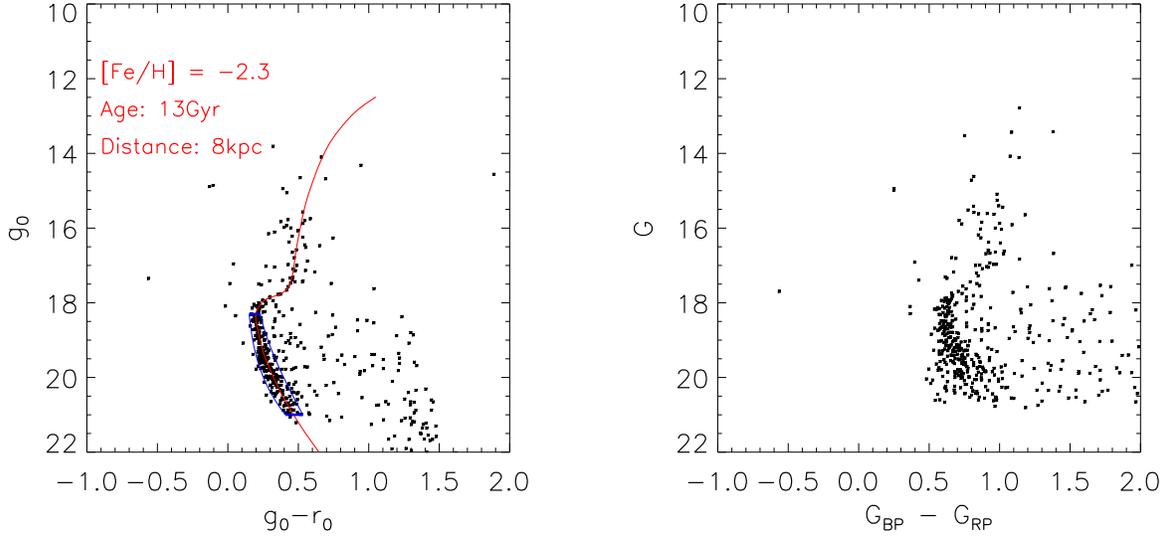}
\caption{The CMDs of GD-1 stars in the red polygon in Fig.~\ref{photo_pm}. Light panel: The 
CMD is from SDSS DR9. The red line is the isochrone with [Fe/H]$ = -2.3$, an age of 13 Gyr 
and a distance of 8 kpc. Right panel: The CMD is from Gaia DR2. The G, G$_{\mathrm{BP}}
$, and G$_\mathrm{RP}$ are the three broad-band magnitudes of Gaia DR2 without 
correction for extinction.
\label{cmd}}

\end{figure}

\begin{figure}
\plotone{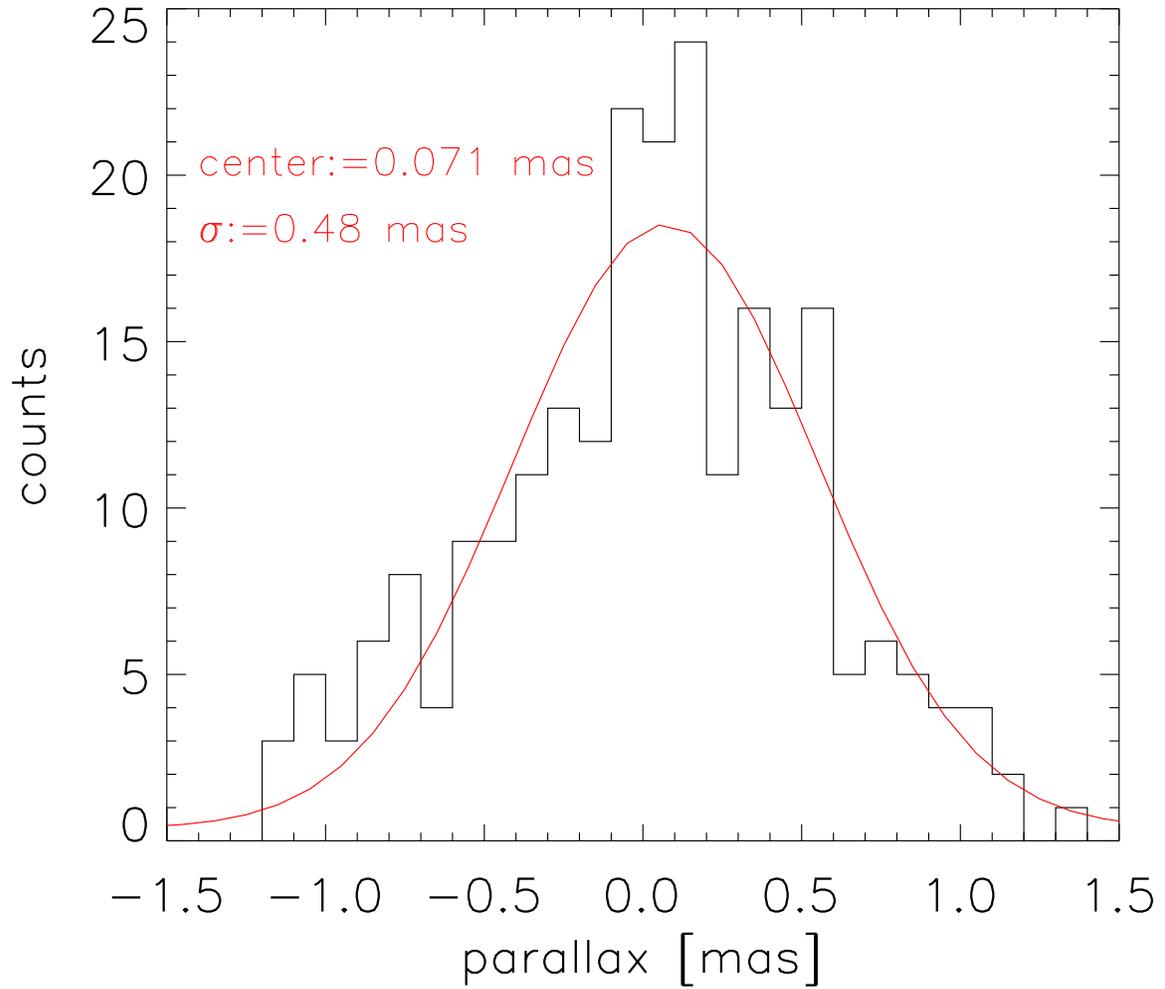}
\caption{The parallax distribution of GD-1 stars enclosed by blue lines in Fig.~\ref{cmd}.
\label{parallax}}

\end{figure}

Fig.~\ref{photo_pm} shows the proper motions of stars in the Gaia DR2 sample within 
$ -60^{\circ} < \phi_1 < -20^{\circ}$ and $-0.2^{\circ} < \phi_2 < 0.2^{\circ}$. The overdensity 
at the bottom left corner clearly stands out. We select the GD-1 candidates with proper 
motions in the red polygon in Fig \ref{photo_pm}. Their CMDs from SDSS DR9 and Gaia 
DR2 are shown in the left and right panels of Fig.~\ref{cmd} respectively. Overlaid in the left 
panel, the red line is the best fit isochrone from  the Dartmouth isochrone library\citep{dot08} 
with [Fe/H] $= -2.3$, an age of 13 Gyr, and a distance of 8 kpc. The area enclosed by the 
blue lines in Fig.~\ref{cmd} is 

\begin{equation}
\left\{
\begin{array}{l}
gr_0 >-36.755917+6.1373654\times g_0 -0.34166210\times g_0^2 +0.0063657131 \times g_0^3 \\
gr_0 < 26.910462 -3.9900350\times g_0 + 0.19378849\times g_0^2 -0.0030286203\times g_0^3\\
18.3 < g_0 < 21
\end{array}
\label{dwarfs}
\right.
\end{equation}  
Stars in this area are dwarfs, whose parallax distribution is shown in Fig.~\ref{parallax}, 
where we can also see that almost all stars have $|\varpi| < 1$ mas. The red line is the fitting
Gaussian function with a mean of $0.071$ mas and a variance of $0.48$ mas. Because of 
the big uncertainty, we cannot estimate the distance of GD-1 stream from the distribution. 
If we apply the parallax correction recommended in \citet{st18}, which is $\sim -0.08$ 
mas, we obtain an estimated typical distance to the stream center from the sun of $\sim 8 \rm 
kpc$.
\par
\begin{figure}
\plotone{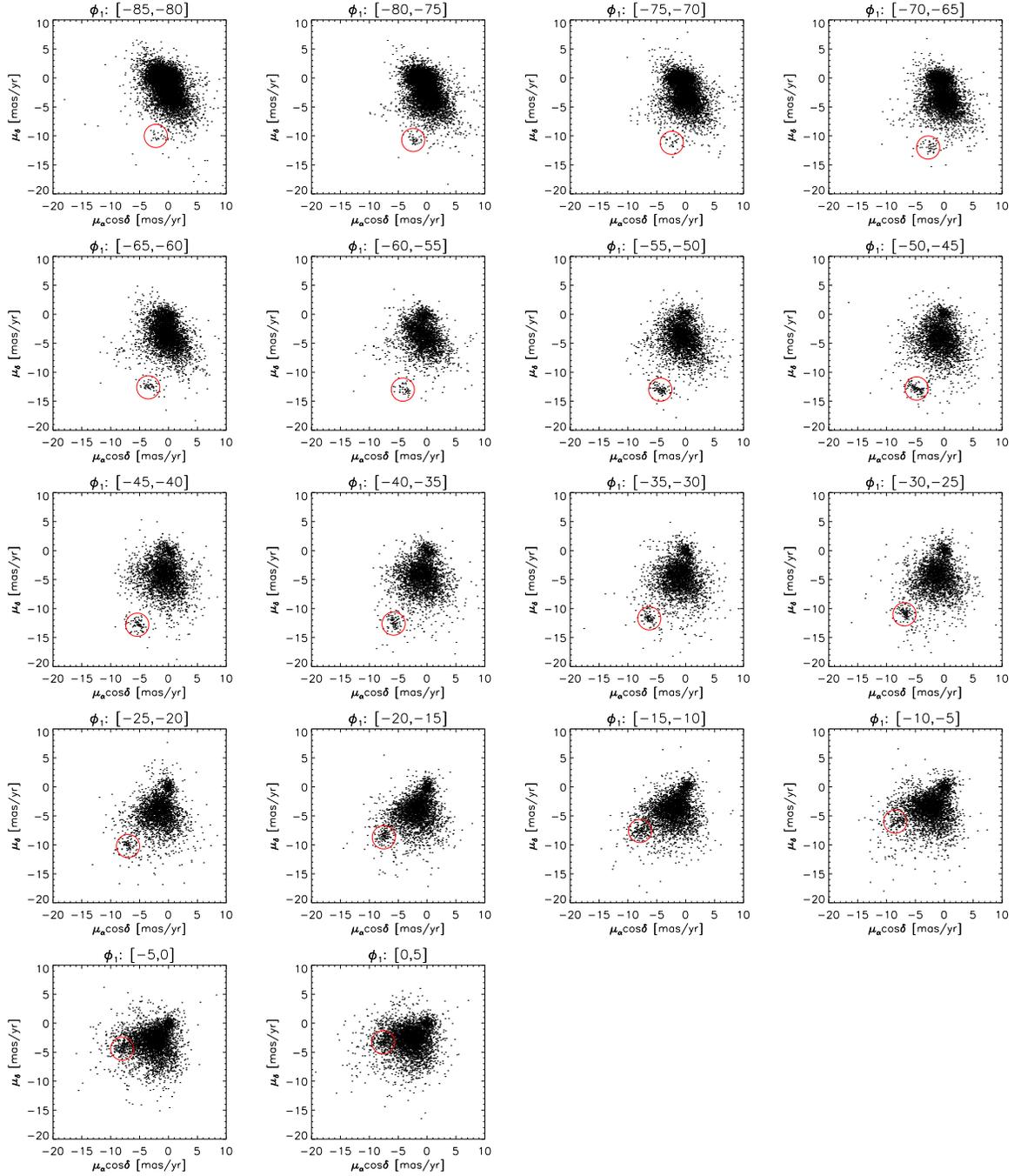}
\caption{The proper motions of stars selected by the CMD of GD-1 dwarf stars with  $|\varpi| 
< 1$ mas,  $-5^{\circ} < \phi_2 < 1^{\circ}$ and distances within $8-10$ kpc. The $\phi_1$ 
range of each 
panel is give in the title. The red circle in each panel is the range where we select GD-1 
candidates. Their centers are given in Table~\ref{region_center}, and radii are all 2 mas/yr. 
\label{gaia_pm}}
\end{figure}

We use the CMD of these dwarfs with  $|\varpi| < 1$ mas and $-5^{\circ} < \phi_2 < 1^{\circ}$ 
to select GD-1 member candidates with distances in 8-10 kpc.
We divide the full range where GD-1 candidate stars are detected $\phi_1: [-85^{\circ}, 
5^{\circ}]$ into 18 regions each with a $5^{\circ}$ interval. Their diagrams of $\mu_{\alpha}
cos\delta -\mu_{\delta}$ are shown in Fig.~\ref{gaia_pm}. The red circle in each panel is 
where we select GD-1 candidates. The radius of each circle is 2 mas/yr, which is large 
enough to cover most GD-1 members, while their fiducial centers are given 
in Table~\ref{region_center}.  
\par
\begin{figure}
\plotone{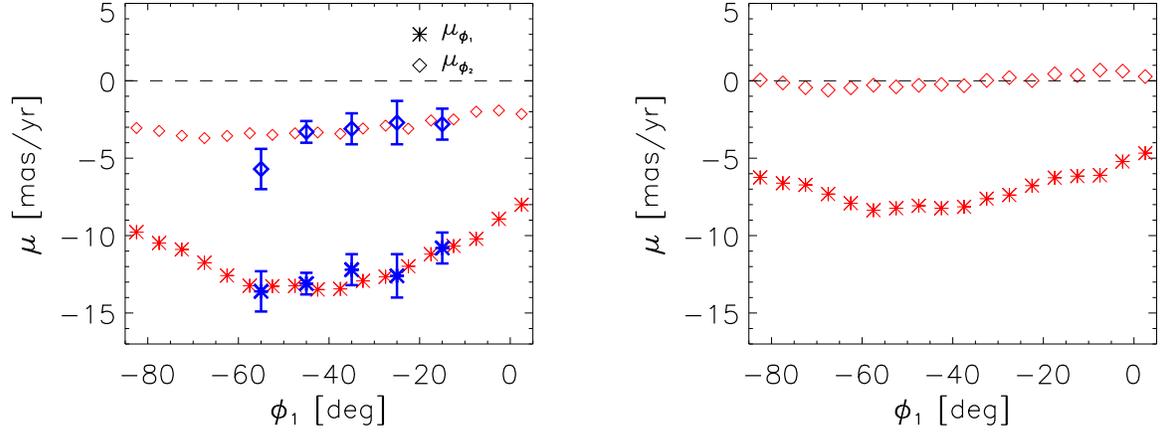}
\caption{The circle centers in Fig.~\ref{gaia_pm}, but have been converted to $(\phi_1, \mu)$ 
coordinate. In each panel, circles are $\mu_{\phi_1}$s, while diamonds are $\mu_{\phi_2}$s, 
and the data from Table~\ref{region_center} are shown in red, while those from Table 4 in 
\citet{kop10} are shown in blue. The proper motions in the right panel have been corrected 
for the Sun's reflex motion, while those in the left panel are not corrected. 
\label{pm_cmp}}
\end{figure}

We select the overdensities in Fig.~\ref{gaia_pm} by hand, then convert the circle 
centers in Fig.~\ref{gaia_pm} to $(\mu_{\phi_1}, \mu_{\phi_2})$ coordinates. We also use the 
correction formula in \citet{kop10} to correct the Sun's reflex motion (note that there is a typo 
in $\mu_{\phi_{1,2,c}} = \mu_{\alpha,\delta}-\mu_{\mathrm{reflex}}$, which should be 
$\mu_{\phi_{1,2,c}} = \mu_{\alpha,\delta} +\mu_{\mathrm{reflex}}$) by assuming the distance 
along $ \phi_1$: $d(\phi_1) = 8$ kpc if $\phi_1 < -20^{\circ}$, $(\phi_1 + 20)\times 0.1$ kpc, 
otherwise, and $V_{\odot} = 220$ km/s. Fig.~\ref{pm_cmp} shows the circle centers in 
Fig.~\ref{gaia_pm} in $(\phi_1, \mu)$ coordinate. In each panel, asterisks are $
\mu_{\phi_1}$s, while diamonds are $\mu_{\phi_2}$s, and these circle centers are shown in 
red, while the data from Table 4 in \citet{kop10} are shown in blue. The proper motions $
\mu_{\phi_1}$ and $\mu_{\phi_2}$ in the right panel have been corrected for the Sun's reflex 
motion, while those in the left panel are not corrected. The left panel shows that these circle 
centers coincide well with data from Table 4 in \citet{kop10}, while the right panel shows that 
$\mu_{\phi_2} \sim 0$ mas/yr, which implies that GD-1 stars move along the GD-1 trace.
\par
\begin{figure}
\plotone{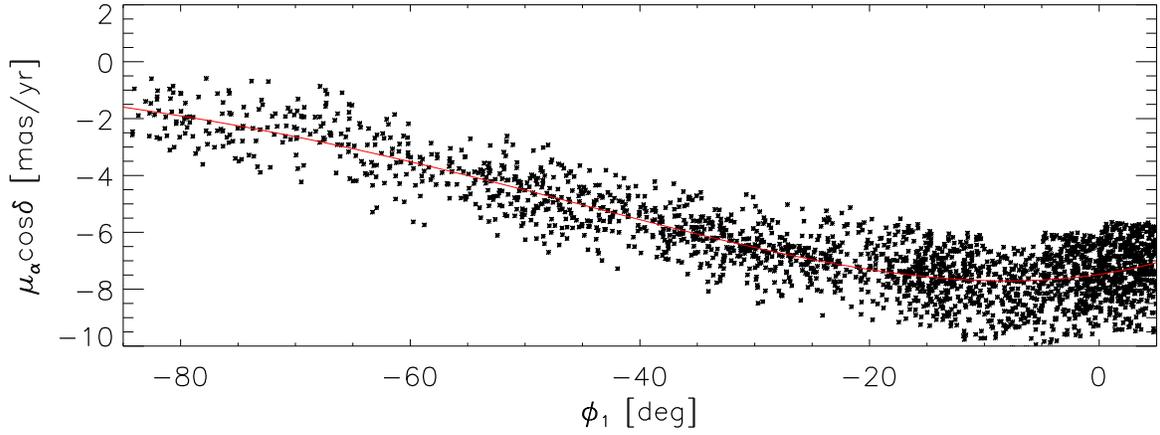}
\caption{ The relationship between Gaia $\mu_{\alpha}cos\delta$ and $\phi_1$. The red line 
is Equation \ref{equ_f1}. \label{phi1_pmra}}
\end{figure}
\begin{figure}
\plotone{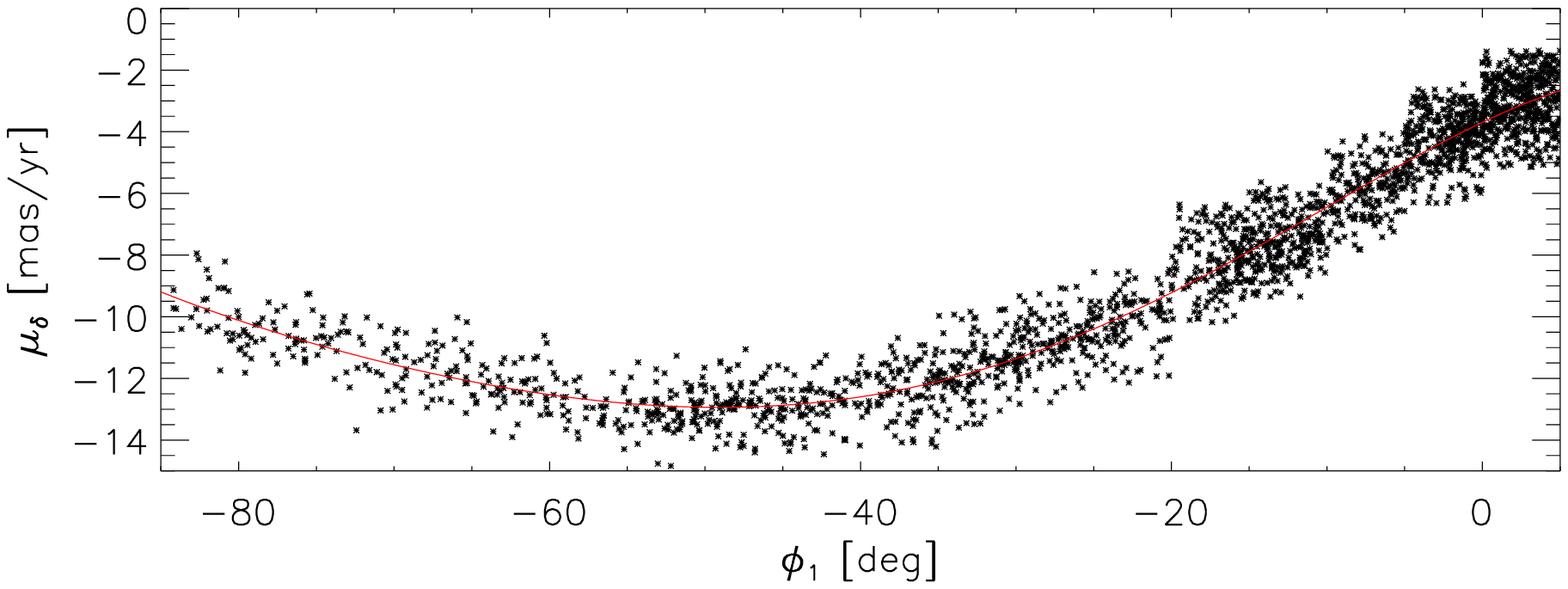}
\caption{ The relationship between Gaia $\mu_{\delta}$ and $\phi_1$. The red line is 
Equation \ref{equ_f2}. \label{phi1_pmdec}}
\end{figure}

\textbf{Fig.~\ref{phi1_pmra} and \ref{phi1_pmdec} show the $\mu_{\alpha}cos\delta$ and 
$\mu_{\delta}$ of the stars in the circles in Fig. \ref{gaia_pm} along $\phi_1$, respectively.
We fit these $\mu_{\alpha}cos\delta$ and $\mu_{\delta}$ by polynomials with the lowest orders that visually go right through the centers of these data along $\phi_1$ in Fig.~\ref{phi1_pmra} and \ref{phi1_pmdec} respectively. The result polynomials are:}
\begin{equation}
\label{equ_f1}
    f_1(\phi_1) = -7.482 + 6.392\times10^{-2}\times \phi_1+ 4.553\times10^{-3}\times \phi_1^2+5.167\times 10^{-5}\times \phi_1^3+1.950\times 10^{-7}\times \phi_1^4
\end{equation}
    and
\begin{equation}    
\label{equ_f2}
    f_2(\phi_1) = -3.698+2.450\times10^{-1}\times\phi_1-5.795\times10^{-3}\times\phi_1^2 -2.550\times10^{-4}\times\phi_1^3 -2.858\times 10^{-6} \times\phi_1^4 -1.121\times 10^{-8}\times\phi_1^5 
\end{equation}  respectively.
The red central line in Fig.~\ref{phi1_pmra} is $f_1(\phi_1)$, while that in 
Fig.~\ref{phi1_pmdec} is $f_2(\phi_1)$. These two equations can 
correct small deviations of the circle centers in Fig.~\ref{gaia_pm} selected by hand.
\par
\begin{figure}
\plotone{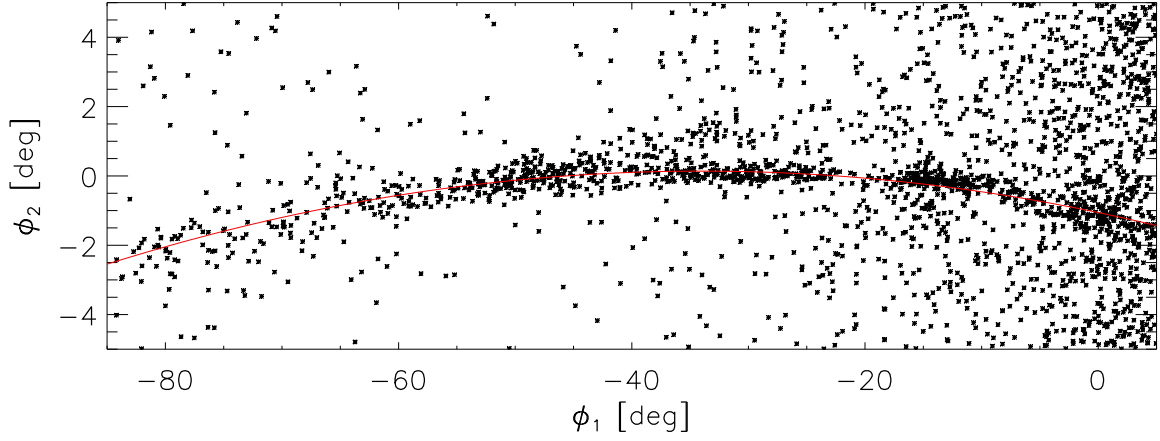}
\caption{ Positions of all stars in the red circles in Fig.~\ref{gaia_pm}. The red line is the 
Equation \ref{equ_f3}. \label{gd1_trace}}
\end{figure}

Now, we map the GD-1 stream using stars by the criteria: (1) They should be in the 
GD-1 dwarf CMD of 8-10kpc, (2) $|\varpi|<1$ mas, and (3) They should be covered by the 
circles in Fig.~\ref{gaia_pm}. Fig.~\ref{gd1_trace}  shows positions of these stars, where the 
GD-1 overdensity clearly stands out and follows the trace. We select some fiducial points 
along the GD-1 trace in $(\phi_1,\phi_2)$, which is listed in Table~\ref{sky_pos}, then fit them 
by a quadratic polynomial:
\begin{equation}    
\label{equ_f3}
    f_3(\phi_1) =  -1.057 - 7.030\times10^{-2} \times\phi_1 -1.033\times10^{-3}\times\phi_1^2
\end{equation}  
, which is the red line in Fig.  \ref{gd1_trace}.  We also introduce
\begin{equation}    
\label{equ_f4}
    \phi_2^{c}(\phi_1) \equiv \phi_2 - f_3(\phi_1)
\end{equation}  
to help us further study the GD-1 stream. 
\par
\begin{figure}
\plotone{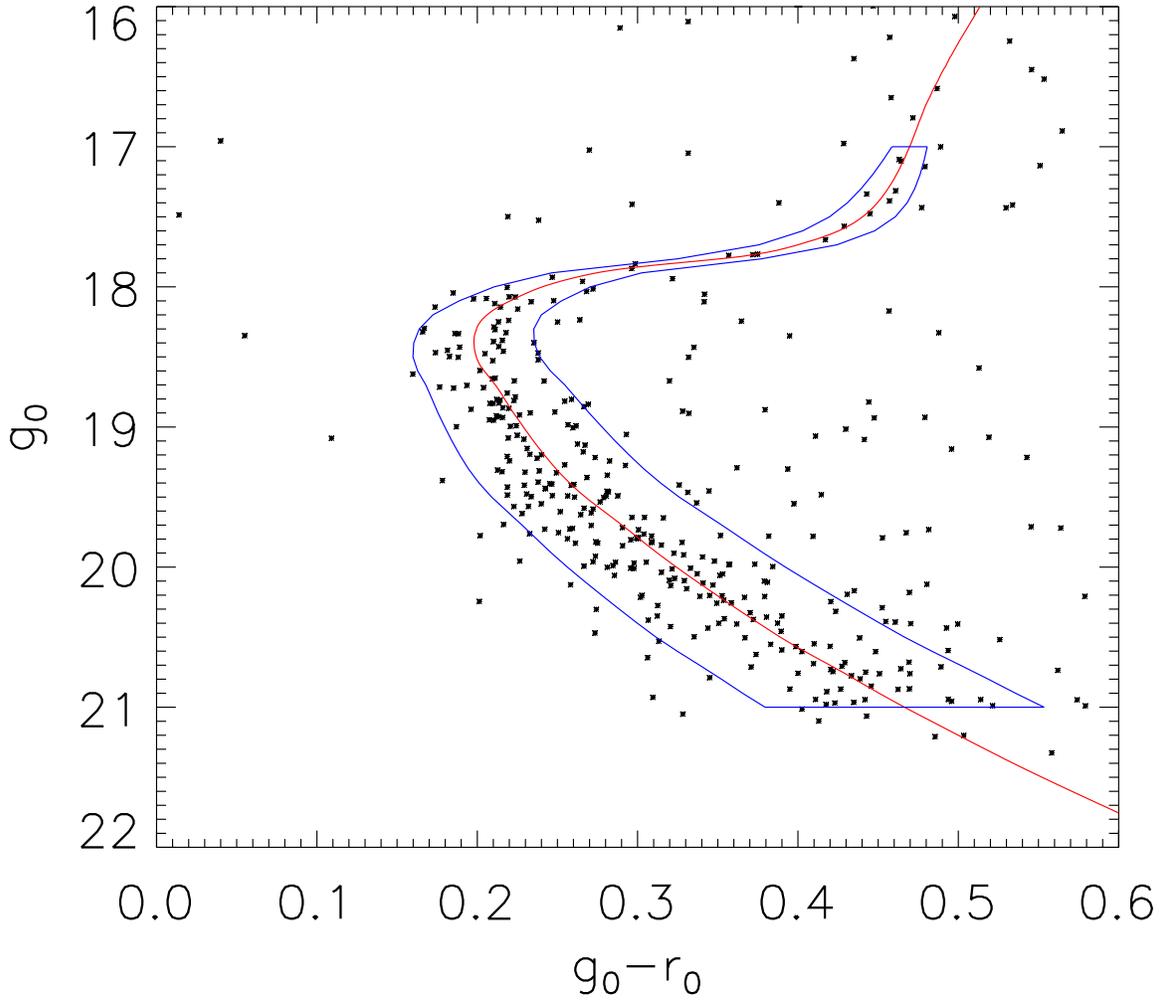}
\caption{ 
The twice $\sigma(g_0)$, which is used to broaden the Dartmouth isochrone between $17 < g_0 < 21$ to generate the template filter, is shown by blue lines. The stars in Fig. \ref{cmd} are shown by black symbols. The red line is the Dartmouth isochrone with [Fe/H]$ = -2.3$, an age of 13 Gyr and a distance of 8 kpc.
 \label{filter}}
\end{figure}

To determine the distances of different parts of the GD-1 stream, we select stars by the 
criteria: (1) $|\varpi|<1$ mas, (2) $|\phi_2^{c}| < 0.5^{\circ}$, (3) $|\mu_{\alpha}cos\delta -
f_2(\phi_1)| <1$ mas/yr and (4) $|\mu_{\delta} - f_3(\phi_1)|<1$ mas/yr. \textbf{The matched 
filter algorithm in \citet{wil09} is used to estimate the distances. We find that the stars of the 
GD-1 stream in Fig. \ref{cmd} within $17 < g_0 < 21$ can be well bounded by $|gr_0(g_0) - 
\mathrm{ISO}(g_0)| < -0.312 + 0.019 \times g_0$ ($\mathrm{ISO}(g_0)$ is the Dartmouth 
isochrone with [Fe/H] = $-2.3$, an age of 13 Gyr and a distance of 8 kpc), which is shown in 
Fig. \ref{filter}, where the bound lines are shown in blue, stars are shown by black symbols 
and the red line is the Dartmouth isochrone with [Fe/H]$ = -2.3$, an age of 13 Gyr and a 
distance of 8 kpc. Thus we use $\sigma(g_0) = 0.5 \times (-0.312 + 0.019 \times g_0)$ for 
Gaussian profiles to broaden the Dartmouth isochrone to generate the template filter at the 
distance of 8 kpc. }
\par
\begin{figure}
\plotone{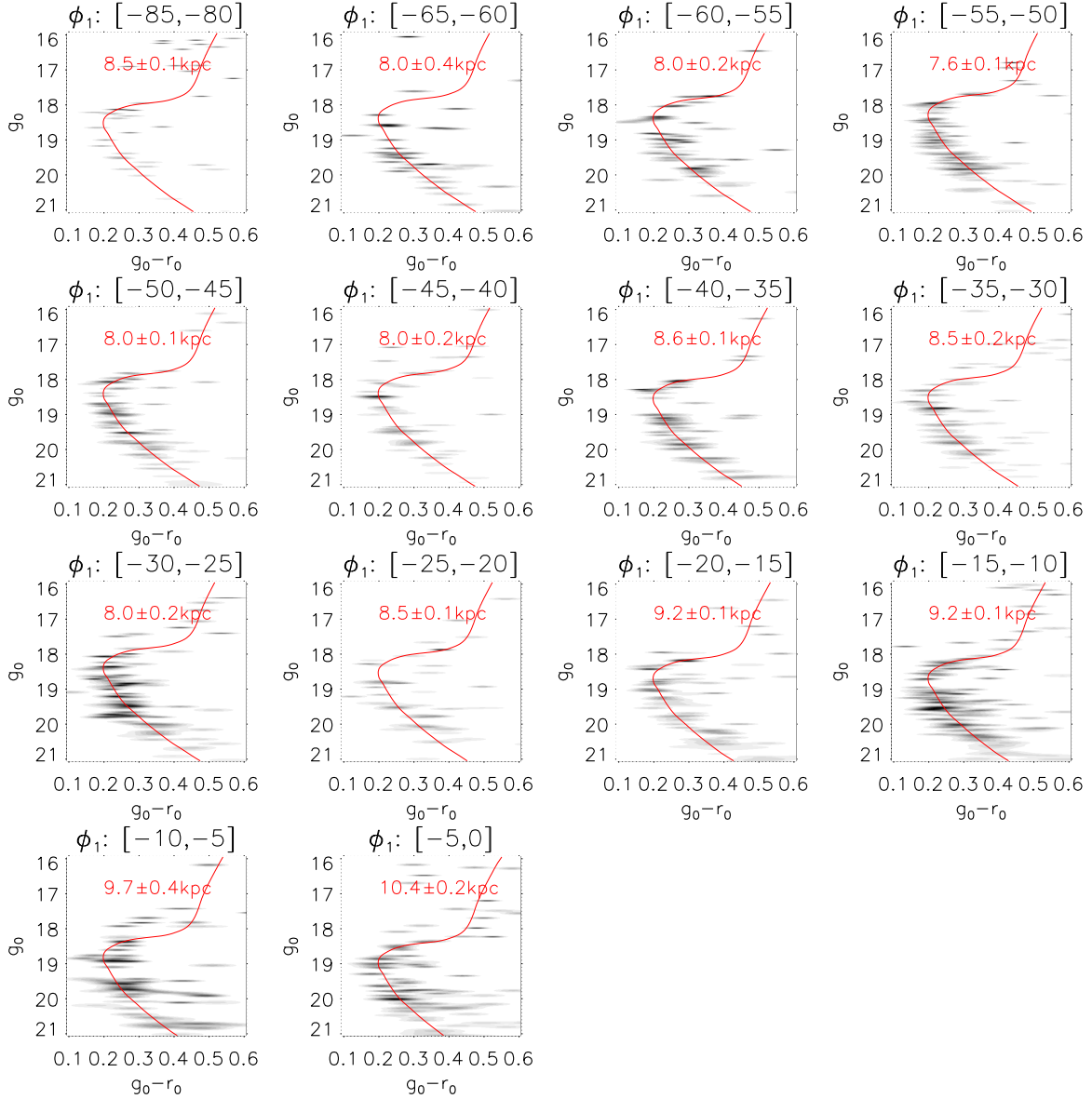}
\caption{ 
The Hess diagrams of 14 regions along the GD-1 trace. In each panel, the red line is the 
isochrone with [Fe/H] $=-2.3$ and an age of 13 Gyr, and the distance with error is given by 
the red number. 
 \label{region_dis}}
\end{figure}

\textbf{We generate Hess diagrams of 18 regions each with a $5^{\circ}$ interval along the $
\sim 90^{\circ}$ GD-1 trace by convolving the SDSS photometric errors on their CMDs. We 
select the Hess diagrams on which the GD-1 dwarfs obviously overwhelm the background 
stars as shown in Fig. \ref{region_dis}. In each Hess diagram, we shift the filter from -0.3 mag 
to 0.9 mag in $g_0$ by the step of 0.01 mag,  then use the Equation 2 - 6 in \citet{wil09} 
with assuming the Hess diagram of the background $\sim 0$ to calculate the distance and 
error of the region (there is a typos in Equation 6 in \citet{wil09}, which should be $
\sigma_{\delta r} = \sqrt{-\frac{\sigma^2(a)}{a(\delta r_m)\frac{d^2a}{d\delta r^2}}}|_{\delta r = 
\delta r_m}$, because the second derivative $\frac{d^2a}{d\delta r^2}$ should be not greater 
than 0 at the maximum). In each panel of Fig. \ref{region_dis}, the red line is a Dartmouth 
isochrone with [Fe/H] = $-2.3$ and an age of 13 Gyr, and the distance with error is shown in 
red. All distances with Gaussian errors are listed in Table~\ref{region_center}.}


\subsection{spectral Data}
\begin{figure}
\plotone{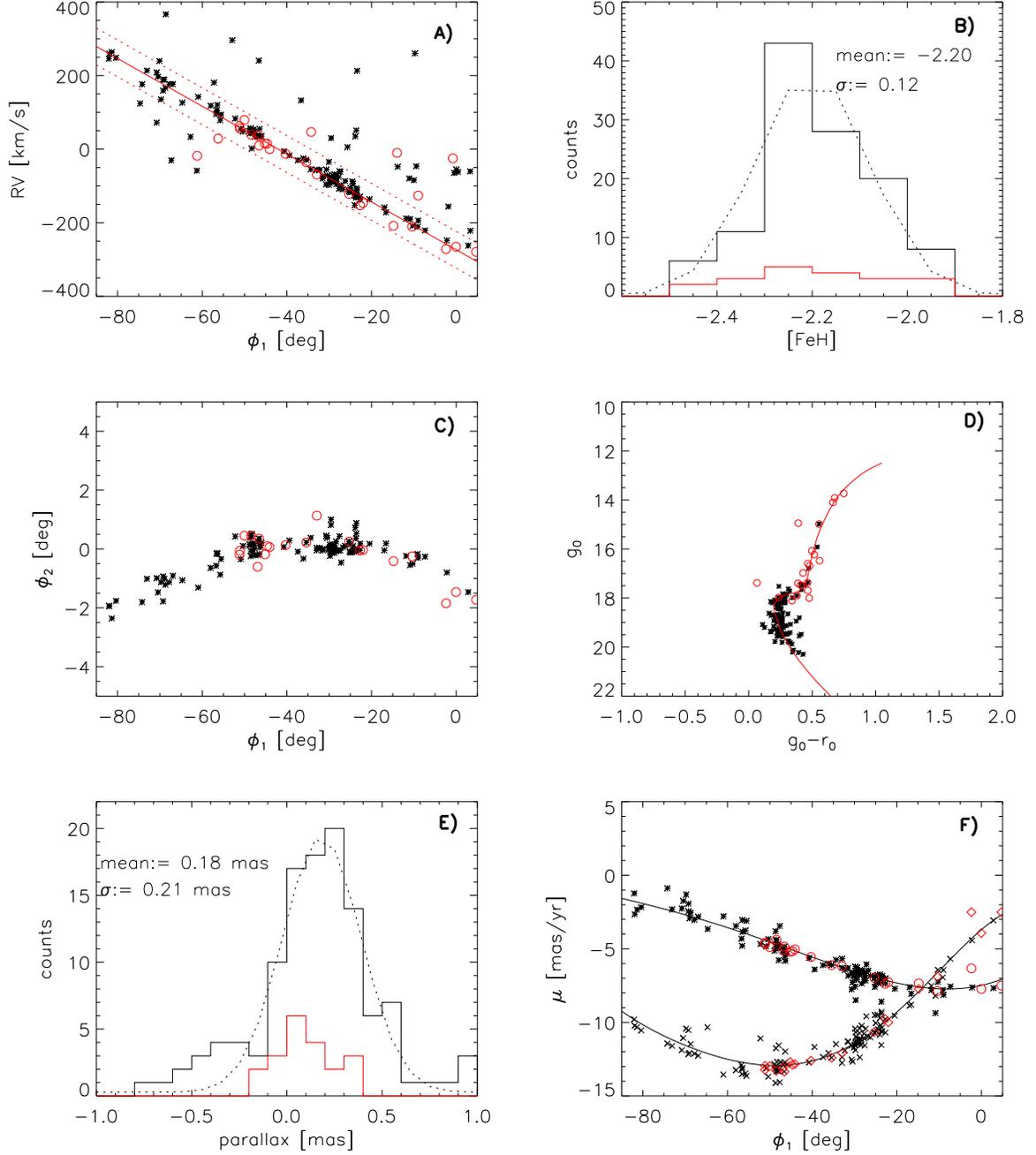}
\caption{The GD-1 candidates in the spectroscopic data of SDSS DR14 and LAMOST DR6 
satisfying the criteria: (1) $|\varpi|<1$ mas, (2)$|\phi_2^{c}| < 1^{\circ}$,  (3) $|\mu_{\alpha}
cos\delta -f_1(\phi_1)| <2$ mas/yr, (4) $|\mu_{\delta} - f_2(\phi_1)|<2$ mas/yr and (5) [Fe/H] 
$<-1.9$ dex. Their radial velocities are shown in panel A. The [Fe/H] distributions, sky 
positions, CMD, parallax distributions and proper motions of the stars within $|RV - 
f_4(\phi_1)| < 50$ km/s are shown in panel B, C, D, E, and F respectively. The red 
symbols and histograms are for the LAMOST stars, while black symbols and histograms are 
for the SDSS stars.
\label{spec}}
\end{figure}

We select the GD-1 candidates in the spectral data of SDSS DR14 and LAMOST DR6 by the 
criteria:  (1) $|\varpi|<1$ mas, (2)$|\phi_2^{c}| < 1^{\circ}$,  (3) $|\mu_{\alpha}cos\delta -
f_1(\phi_1)| <2$ mas/yr, (4) $|\mu_{\delta} - f_2(\phi_1)|<2$ mas/yr and (5) [Fe/H] $<-1.9$ 
dex. For a star that have multiple spectra and multiple measurements, its parameters with 
errors are the mean values of these measurements. In panel A of Fig.~\ref{spec}, the spectral 
radial velocities of SDSS DR14 along $\phi_1$ are shown by black asterisks, while those of 
LAMOST DR6 are shown by red circles, and the central red line is  
\begin{equation}    
\label{equ_f5}
    f_4(\phi_1) =  - 273.4 -6.5 \times \phi_1
\end{equation}  
km/s, while the dotted lines are $f_4(\phi_1)\pm50$ km/s. We can see that stars are crowded 
in this range, where GD-1 members are located. The spectroscopic stars in SDSS DR14 
and LAMOST DR6 that meet all of the above four criteria and also have radial velocities $|RV 
- f_4(\phi_1)| < 50$ km/s are selected as GD-1 spectroscopic stream member candidates. 
One hundred and thirty-six spectra for 116 individual stars from SDSS DR14 and 32 spectra 
for 20 individual stars from LAMOST DR6 are given in Table~\ref{spec_sdss} and 
Table~\ref{spec_lamost} respectively.
\par
Panel B of Fig.~\ref{spec} shows the [Fe/H] distributions of these spectroscopic member 
candidates. The black histogram is for SDSS stars, which is fitted by a Gaussian function and 
indicated by a dotted profile, while the red histogram is for LAMOST stars. The fitted value of 
[Fe/H] histogram  for SDSS stars is $-2.20\pm 0.12$ dex, while the peaks of [Fe/H] 
histograms for SDSS and LAMOST stars  are all at $ \sim -2.25$ dex, which are well 
consistent with $-2.3$ dex that the photometric data give in 
Fig.~\ref{cmd}.   \citet{gao15} has shown that for low metal stars, [Fe/H] (\citep{wu14} less 
than $-1.5$ dex in LAMOST DR1, which are determined by LASP with ELODIE interpolator) 
are systematically measured higher when compared with those in the PASTEL 
catalog\citep{sou10}.  Thus the actual intrinsic metallicity of the GD-1 stream we estimate no 
more than $-2.20$ dex, with an estimated error of about 0.12 dex.
\par
Panel C of Fig.~\ref{spec} shows the sky positions of these candidates, while their CMD 
is shown in panel D. In these two panels, SDSS stars are indicated by black asterisks, 
while LAMOST stars are indicated by red circles. The red line in panel D is the Dartmouth 
isochrone with [Fe/H] = $-2.3$, an age of 13 Gyr, and a distance of 8 kpc. The former panel 
shows that SDSS stars are distributed along the whole GD-1 trace, while LAMOST stars are 
only extended to $\phi_1 \sim -52^{\circ}$ from  $\phi_1 \sim 5^{\circ}$; The latter panel 
shows most LAMOST stars are giants or subgiants, while SDSS stars are mostly G dwarfs 
and F turnoff stars.
\par
The parallax distributions of these spectroscopic candidates from SDSS and LAMOST 
catalogues are shown in panel E by black and red histograms respectively. The parallax 
distribution of SDSS is fitted by a Gaussian function, which is shown by a dotted profile, with 
a mean of 0.18 mas and a variance of 0.21 mas. Panel F shows the proper motions of these 
spectroscopic candidates. The $\mu_{\alpha}cos\delta$ and $\mu_{\delta}$  of SDSS 
spectroscopic candidates are shown by black asterisks and black crosses respectively, while 
those of LAMOST spectroscopic candidates are shown by red circles and red diamonds 
respectively. Equation \ref{equ_f1} and \ref{equ_f2} are overplotted by black lines as 
guidelines.

\section{Orbit fitting}\label{sec:orbit}
Now we have 6D phase space information along the GD-1 trace: (1) radial velocities from 
Table~\ref{spec_sdss}, \ref{spec_lamost} and Table 1 in \citet{kop10}, (2, 3) proper motions 
from Equation \ref{equ_f1} and \ref{equ_f2}, (the errors in $\phi_1 : [-60^{\circ}, -10^{\circ}]$ 
are less than 0.4 mas/yr, beyond this range the errors are harder to estimated, but we estimate
a large upper bound of 1 mas/yr ), (4) the isochrone fitting distances from 
Table~\ref{region_center} and (5, 6) the sky positions from Table~\ref{sky_pos}.  We use 
three Galactic potential models from \textit{galpy}\footnote{\url{http://github.com/jobovy/
galpy}}\citep{bov15} to fit the GD-1 trace: Model 1 is a spherical \textit{MWPotential2014}, 
Model 2 is \textit{LogarithmicHaloPotential} with the potential flattening $q_{\Phi} =0.9$,  the 
distance from the sun to the Galactic center $ro=8.0$ kpc and  the circular velocity at sun 
$vo=220$ km/s, and Model 3 is similar to Model 2 but $ro=8.5$ kpc.
\par
\begin{figure}
\plotone{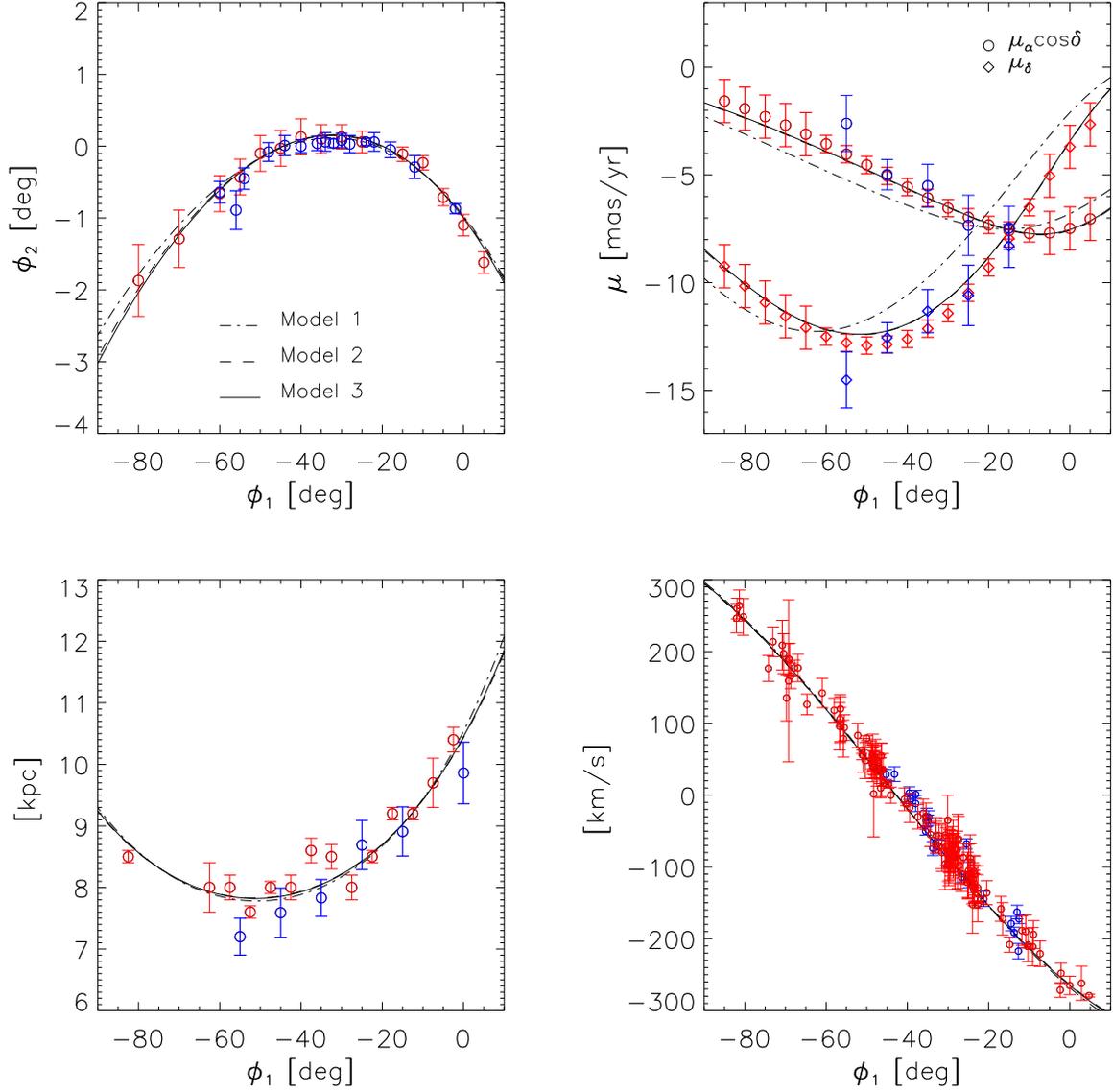}
\caption{ 
The fitted orbits by three models. The red data are from this paper, while the blue data are 
from \citet{kop10}. The red data and blue radial velocities are used to fit the GD-1 orbits by 
models, while other blue data are used for comparison. The dashed dotted, dashed and solid 
lines are the fitted orbits from Model 1, 2 and 3 respectively.
 \label{orbit_fit}}
\end{figure}

We construct an objective function of how well each model fits based on 6D information. 
The objective function is defined as the sum of $\chi^2$ of each dimension divided by the 
number of data points (data of five of the six dimensions are functions of $\phi_1$). The $
\chi^2$ of each dimension is the sum of squares of differences between data and modal 
divided by the data errors. We minimize the value of the objective function, then obtain the 
GD-1 orbit under a given modal. Results are shown in Fig.~\ref{orbit_fit}.  The red data are 
from this paper, while the blue data are from \citet{kop10}. The red data and blue radial 
velocities are used to fit the  GD-1 orbit by models, while other blue data are used for 
comparison. The dashed dotted, dashed and solid lines are the fitted orbits from Model 1, 2 
and 3 respectively. From this figure, and the extended, added data gather here, we find the 
data can be well fitted by Models 2 and 3, while Model 1 has a much poorer fit. The orbit from Model 3 
shows that the stream has an eccentricity $e \sim 0.3$, perigalacticon of 14.2 kpc, 
apogalacticon of 27.0 kpc and inclination approximately $i \sim 40^{\circ}$.


\section{Discussion}
\label{sec:discuss}
In this section, we continue to inspect the GD-1 stream and its environs.
We select stars by criteria: 
(1) $|\varpi|<1$ mas, (2) $|\mu_{\alpha}cos\delta -f_1(\phi_1)| <2$ mas/yr, (3) $|\mu_{\delta} - 
f_2(\phi_1)|<2$ mas/yr, (4) $-85^{\circ}< \phi_1 < 5^{\circ}$, and (5) $-0.4 < gr_0 < 1.1$. 
We use 
\begin{equation}    
\label{equ_f6}
  D(\phi_1) = 10.4 + 0.105 \times \phi_1 + 0.00103 \times \phi_1^2
\end{equation} 
(in kpc) to fit the distance given by Modal 3 in Section \ref{sec:orbit} along the GD-1 
trace, with error less than 0.1 kpc. Then we shift all stars to 8 kpc by distance modulus $DM 
= 5\lg(D(\phi_1)/8)$ and denote $g_0^c = g_0 - DM$. We also obtain the absolute magnitude 
of SDSS $g$:  $M_g = g_0 - 5\lg(D(\phi_1)) - 10$.  We will use the refined magnitude 
definition here in the section that follows to explore density
variation along the stream.
\subsection{Stellar Density Fluctuation along the GD-1 Trace}
\begin{figure}
\plotone{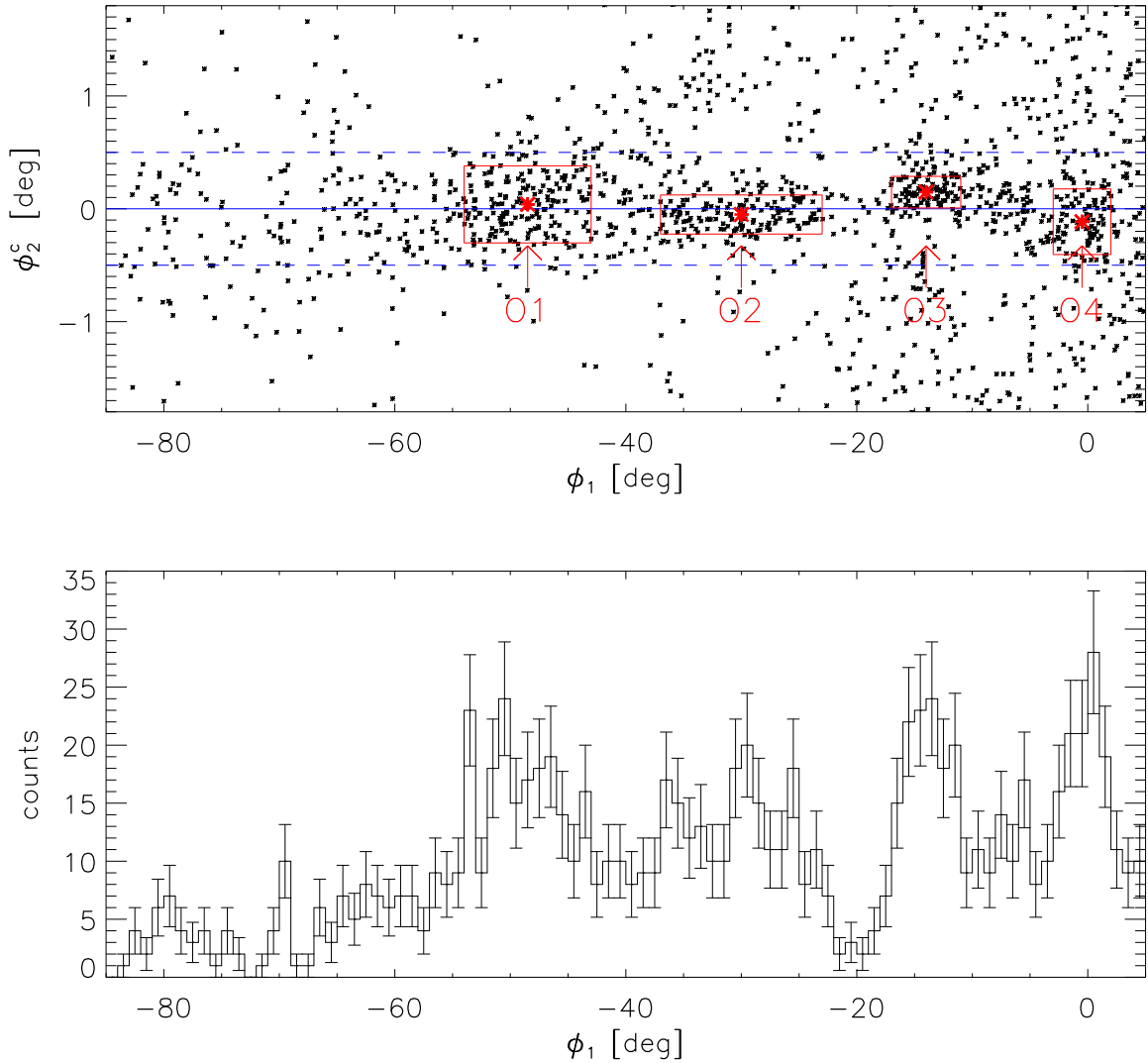}
\caption{ Upper panel: The sky positions of stars selected by Equation \ref{dwarfs}, where 
$g_0$ is replaced by $g_0^c$. The blue line is $\phi_2^c = 0^{\circ}$, while the blue dotted 
lines are $\phi_2^c = \pm 0.5^{\circ}$. Bottom panel: The stellar density histogram with Poisson error along the GD-1 trace within two blue dashed lines in the upper panel. 
 \label{dens}}
\end{figure}

In this refined CMD (in $(gr_0, g_0^c)$ coordinates), we use Equation \ref{dwarfs} to select 
GD-1 dwarfs, where $g_0$ is replaced by $g_0^c$. The sky positions of these dwarfs are 
shown in the upper panel of Fig.~\ref{dens}, where the blue line is $\phi_2^c = 0^{\circ}$ and 
the blue dotted lines are $\phi_2^c = \pm 0.5^{\circ}$. The stellar counts with Poisson error 
along GD-1 trace between the two blue dotted lines are shown in the bottom panel. 
While we acknowledge that because we use Gaia DR2 to select candidate stream members, 
and that catalog is limited to $g \sim 21$, that the statistical significance of suspect over or 
underdensities may also be limited, we do find that:

\par
(1) There are four candidate overdensities along the GD-1 trace. These four overdensities are 
at $\phi_1: [-54^{\circ}, -43^{\circ}]$, $[-37^{\circ}, -23^{\circ}]$, $[-17^{\circ}, -11^{\circ}]$ and 
$[-3^{\circ}, 2^{\circ}]$, which are denoted as O1, O2, O3 and O4 respectively; 
\par
(2) There is a gap centered around $\phi_1 \sim -21^{\circ}$, which has no GD-1 candidate 
member star;
\par
(3) There are two candidate underdensitiies at $\phi_1 \sim -40^{\circ}$ and $\phi_1 \sim 
-8^{\circ}$;
\par
(4) The width of stream visually broadens along the GD-1 trace from O3 along two directions;
\par
(5) There is a noticeable wobble in O1;
\par
(6) The stellar density drops suddenly to background level at $\phi_1 \sim 2^{\circ}$;
\par
(7) A spur originates at  $\phi_1 \sim -40^{\circ}$, where the O1 and O2 are separated, which 
is discussed by \citet{pri18} and also shown in Figure 10 of \citet{kop10}. We conjecture that 
it results from GD-1 stars pulled out of the stream by an encounter. Thus, O1 and O2 
were once on a single overdensity, but separated later.
\par
\begin{figure}
\plotone{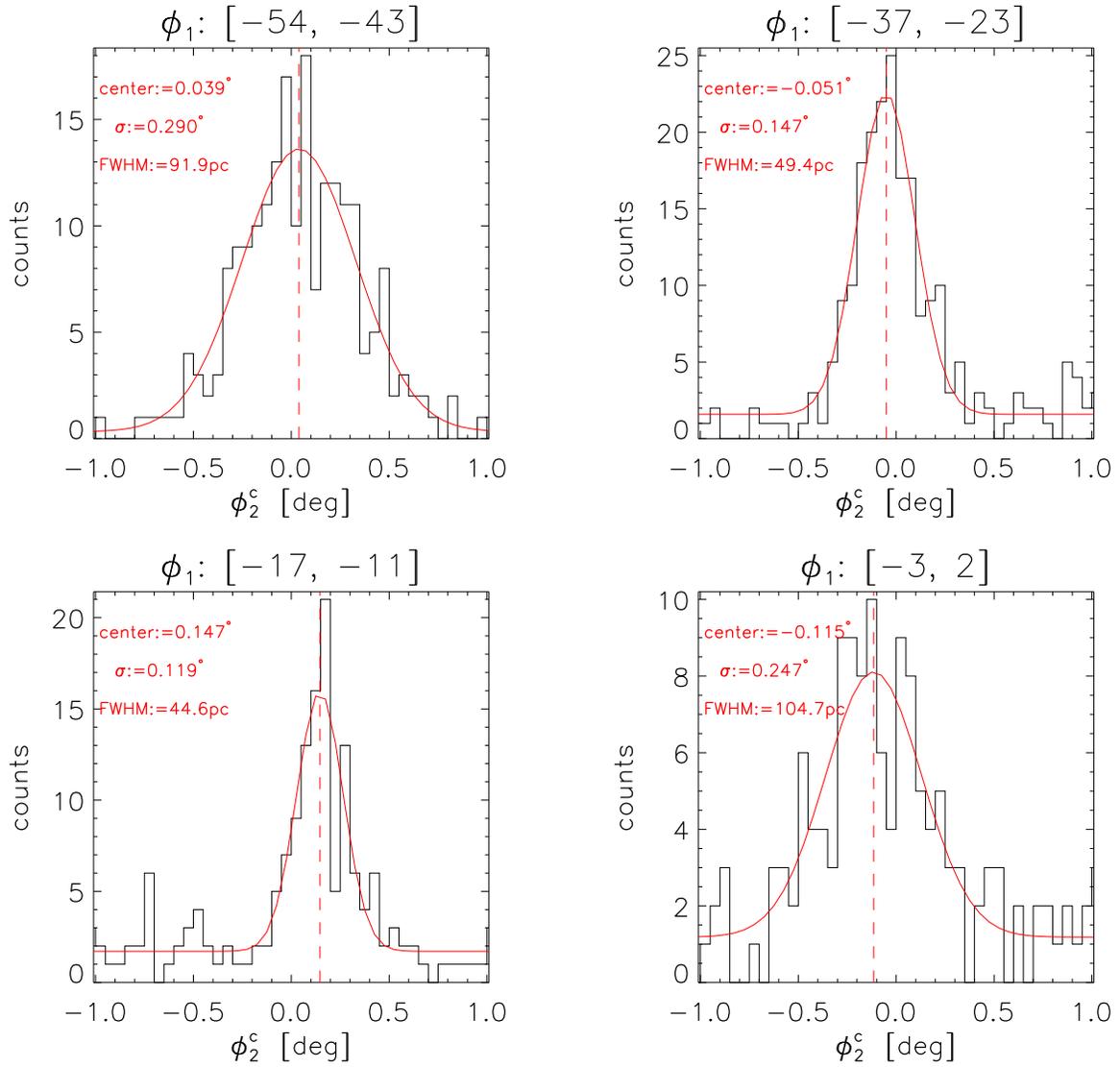}
\caption{ 
Stellar density profiles for four overdensities. The fitted Gaussian functions are overplotted 
on the histograms by red lines, and the fitted centers and dispersions are also given.
 \label{dens_prof}}
\end{figure}

Fig.~\ref{dens_prof} shows the density profiles of four overdensities O1, O2, O3 and 
O4 along $\phi_2^c$. We fit these profiles by Gaussian functions, and in each panel, the 
Gaussian width $\sigma$, center and full width at half maximums(FWHM) are also given in 
red. Their lengths along $\phi_1$ and FWHMs in degrees along the $\phi_2^c$ are shown by 
the red rectangles in the upper panel of Figure \ref{dens}. All information about these four 
overdensities are given in Table \ref{gd1_od}, where the areal densities of dwarfs are the 
ratios of dwarf number(calculated from Gaussian fitting) divided by areas of rectangles in the 
upper panel of Fig. \ref{dens}. 
Their Gaussian centers are $0.039^{\circ}, -0.051^{\circ}, 0.147^{\circ}$ and 
$-0.115^{\circ}$ in $\phi_2^c$ respectively, while Gaussian widths are $0.290^{\circ}, 
0.147^{\circ}, 0.119^{\circ}$ and $0.247^{\circ}$ respectively, which correspond to FWHMs of 
91.9 pc, 49.4 pc, 44.6 pc and 104.7 pc respectively by the distance of Equation \ref{equ_f6}. 
These FWHMs  show that the narrowest place of the GD-1 stream is in O3, and then 
broadens gradually along two directions. Besides, O3 has the highest areal density of 
dwarfs as shown in Table \ref{gd1_od}. For stars stripped from the progenitor cluster earlier 
would move away further both in position and velocity from the progenitor\citep{bov14}. So 
O3 centered $(\phi_1, \phi_2^c) \sim (-14^{\circ}, 0.15^{\circ})$($(\alpha, \delta)\sim 
(178.18^{\circ}, 52.19^{\circ})$), which has the narrowest stream width, highest stellar areal 
density and also shows a noticeable wobble, is the most likely candidate for the GD-1 stream 
progenitor.
 \par
Systems at the end of their bound lifetime like GD-1 are not expected to be stripped 
smoothly, so at least some fluctuations from the normal trace are expected to simply be the 
result of the last few stripping episodes being stochastic and erratic. But given the overall 
density structure, it seems possible that the local density is affected by the encounters 
that created the under dense features. After all, the energy of the stars is redistributed, and 
shocks and caustics could mess up the smooth stream density.   A more detailed dynamical 
model of the stream over time including encounters with dense structures and tracing the 
stream as it passes through the plane of the disk will help determine the nature of these 
encounters.

\subsection{Blue Horizontal Branch(BHB) Stars and Blue Stragglers(BS)}
\label{subsec:bhb}
\begin{figure}
\plotone{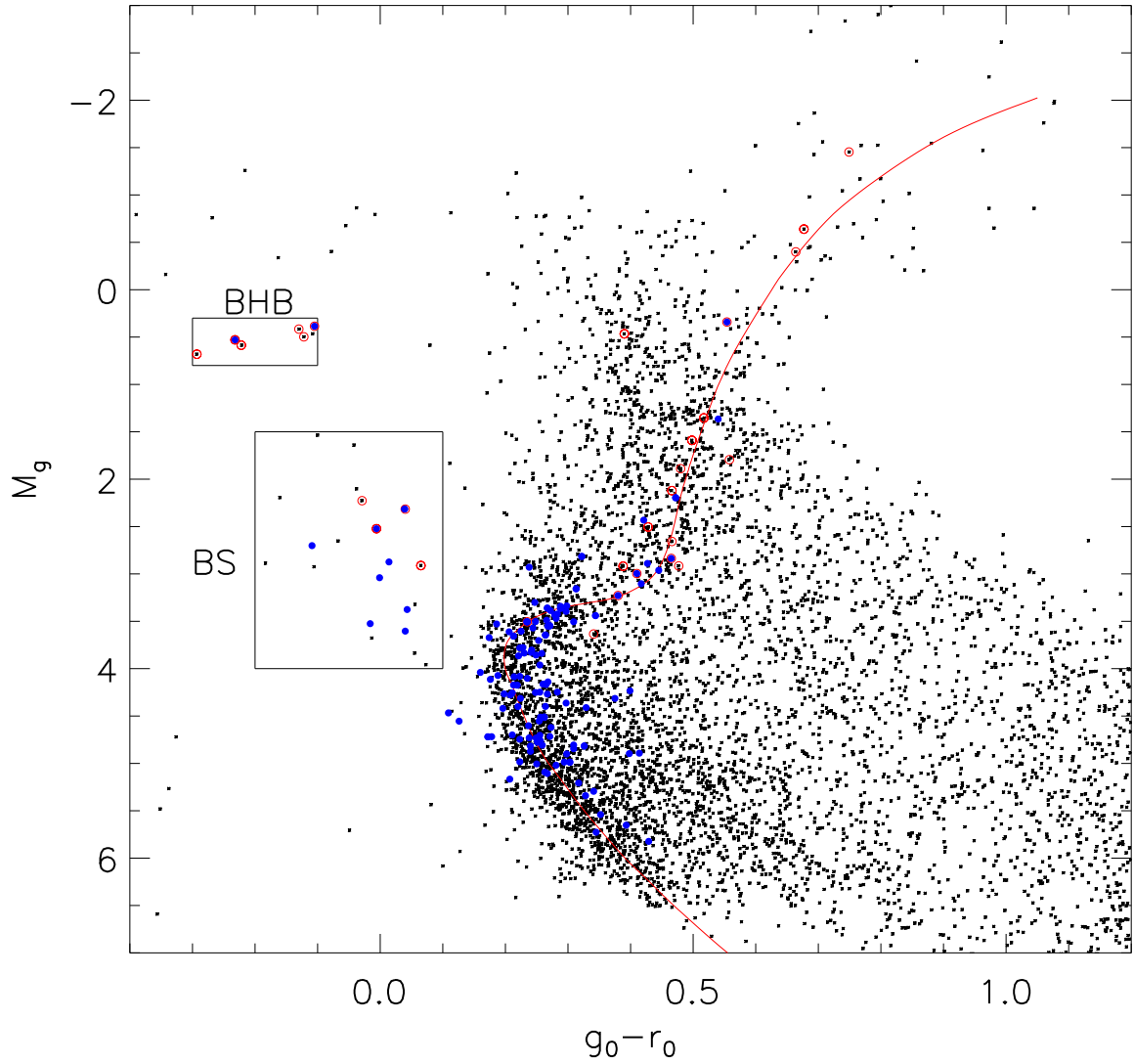}
\caption{Stars are selected by the criteria in Section \ref{subsec:bhb}. The red curve is 
the isochrone with [Fe/H] $ = -2.3$, and an age of 13 Gyr. BHB stars 
and blue stragglers are shown in rectangles. The spectroscopic stars from SDSS DR14 and LAMOST DR6 are shown by red and blue circles respectively.
 \label{gd1_cmd_all}}
\end{figure}

In this subsection, we require selected stars satisfying the following criteria: (1) $|\varpi| < 1$ 
mas; (2) $-85^{\circ} < \phi_1 < 5^{\circ}$; (3) $| \mu_{\alpha}cos\delta - f_1(\phi_1)| < 2$ 
mas/yr; (4) $|\mu_{\delta}-f_2(\phi_1)| <2$ mas/yr; (5) $-1^{\circ} < \phi_2^c < 1^{\circ}$. Their 
CMD is shown in Fig.~\ref{gd1_cmd_all}. 

\begin{figure}
\plotone{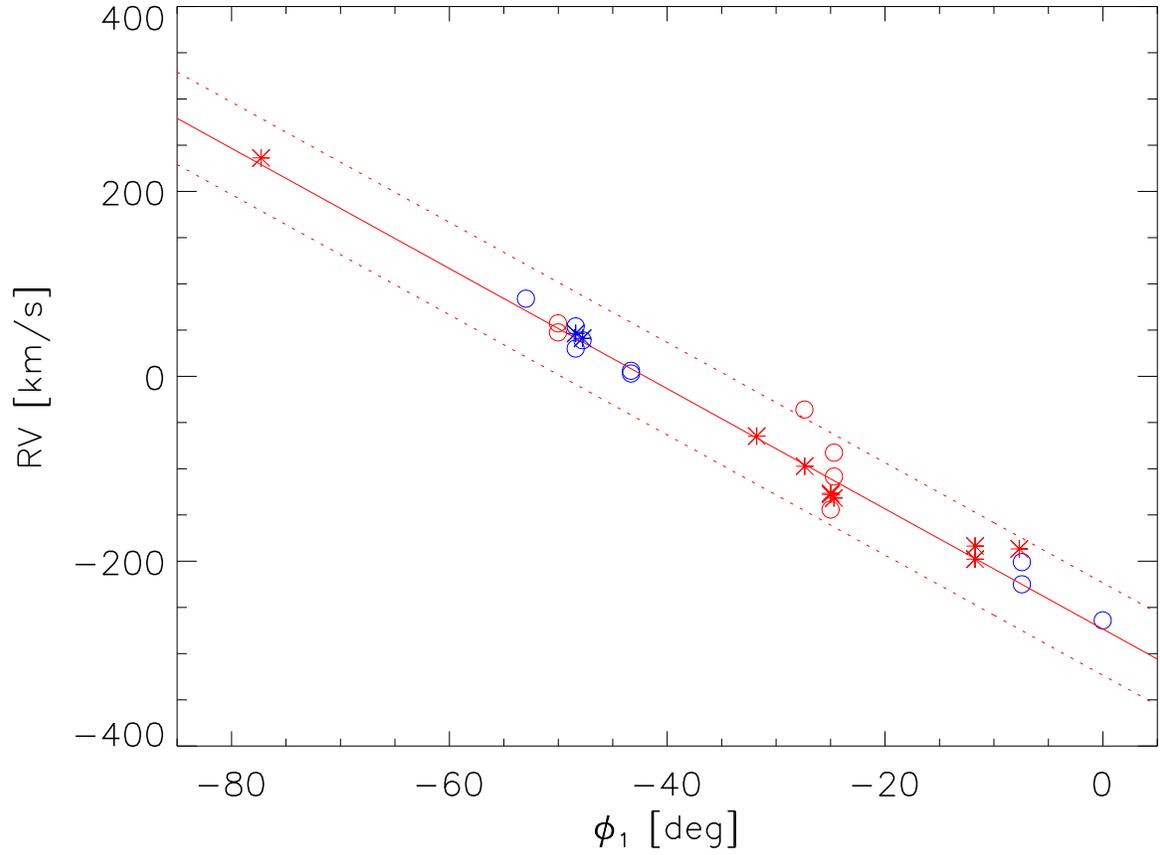}
\caption{ 
The blue and red symbols are the BHB and BS stars respectively. Circles are the stars from 
LAMOST DR6, while asterisks are the stars from SDSS DR14. The red circle far from the 
central line results from low signal-to-noise of LAMOST spectrum. The central red line is 
$f_4(\phi_1)$ km/s, while the two dotted lines are $f_4(\phi_1)\pm 50$ km/s. One symbol 
represent one spectrum, so a star may have several symbols.
 \label{gd1_blue_spec_rv}}
\end{figure}

Because the spectroscopic parameters for hot stars are more difficult to derive, due to less 
characteristic spectral lines, different physical mechanism and sparse templates, we only give 
radial velocities for stars with $gr_0 < 0.1$ from LAMOST DR6 and SDSS DR14, which are 
shown in Fig. \ref{gd1_blue_spec_rv}, where one symbol represent one spectrum, so a star 
may have several symbols. From the figure we can see that the radial velocities of all blue 
stars are well within the radial velocity range of GD-1 stream, so we believe all these 
spectroscopically observed stars are GD-1 members. All blue stars with $gr_0 < 0.1$ from 
spectroscopic catalogues of LAMOST DR6 and SDSS DR14 are listed in 
Table~\ref{gd1_blue_spec_lamost} and Table~\ref{gd1_blue_spec_sdss} respectively.
\par
We overplot all spectroscopic GD1 member stars from LAMOST DR6 and SDSS DR14 by 
red and blue circles in Fig.~\ref{gd1_cmd_all} respectively -- the positions of BHB and BS 
stars of the GD-1 stream are clearly shown. We select BHB stars by the criteria of 
$0.3<M_g<0.8$ and $ -0.3 < gr_0<-0.1$, and BS stars by the criteria of $1.5<M_g<4$ and  $ 
-0.2 < gr_0<0.1$. There are total seven BHB and 21 BS stars of the GD-1 stream, which are 
listed in Table~\ref{gd1_blue_photo}. In LAMOST DR6 and SDSS DR14, there are six and 
two spectroscopic BHB stars respectively, and four and eight spectroscopic BS stars 
respectively. Not counting spectra of the same star, we have a total of six new spectroscopic 
BHB and ten new spectroscopic BS candidate stream members.

\section{Conclusion} \label{sec:con}
In this paper, combining spectroscopic and photometric data from SDSS DR14, high 
precision astrometric data from Gaia DR2 and spectroscopic data from LAMOST DR6, add 
significantly to the numbers of GD-1 stream high confidence candidate members and obtain a 
full 6D phase-space map of GD-1 stream with high precision. We conclude this information 
about the GD-1 steam and its stellar contents:
\par
(1) The GD-1 stream is the relic of a very old, low dispersion object, such as a metal-poor 
globular cluster or possibly an ultra-faint dwarf galaxy, with  [Fe/H] $\sim -2.3$. \par
(2) Its observed length today on the sky from the sun is at least $90^{\circ}$.\par
(3) There are four candidate stellar overdensities, one candidate gap and two candidate 
stellar underdensities along the stream trace. The trace appears to break at $\phi_1 = 
2^{\circ}$.\par 
(4) The GD-1 progenitor is most likely located at  $(\phi_1, \phi_2^c) \sim (-14^{\circ}, 
0.15^{\circ})$(i.e. $(\alpha, \delta)\sim (178.18^{\circ}, 52.19^{\circ})$).
\par
(5) A spur originating about $\phi_1 \sim -40^{\circ}$ seems to consist stream stars pulled off 
from the GD-1 stream by an encounter with a massive object.\par
(6) We find six new BHB and 10 BS spectroscopic members of GD-1.
\par
In summary, thanks to Gaia DR2, we can obtain much detailed phase space information 
about the GD-1 stream, which can help us to explore the nature of GD-1 stream and study 
the Galactic potential. The spectra of stream members from SDSS DR14 and LAMOST DR6 
helps to understand the origin of the progenitor of the GD-1 stream. Four apparent 
overdensities, the gap, the spur, and other stellar fluctuations along the stream can help us to 
understand the history of GD-1 disruption process and possibly the nature and frequency of 
massive dark matter subhalo encounters.
    
\begin{table}\centering
\caption{Circle centers in Fig.~\ref{gaia_pm} and distances in Fig.~\ref{region_dis}}
	\label{region_center}

	\begin{tabular}{ccc}
		\hline
		$\phi_1$ range &  $(\mu_{\alpha}cos\delta, \mu_{\delta})$ & distance\\
		$[$deg] & [mas/yr] & [kpc] \\\hline
		$[-85, -80]$ & $(-2.2, -10.0)$ & $8.5\pm0.1$\\
       $[-80, -75] $& $(-2.4, -10.7)$&\\
$ [-75, -70] $& $(-2.4, -11.2)$&\\
$ [-70, -65] $& $(-2.8, -12.0)$&\\
$ [-65, -60] $& $(-3.5, -12.6)$ & $8.0 \pm0.4$ \\
$ [-60, -55] $& $(-4.2, -13.0)$ & $8.0 \pm 0.2$\\
$ [-55, -50] $& $(-4.4, -13.0)$& $7.6 \pm 0.1$\\
$ [-50, -45] $& $(-4.8, -12.8)$ & $8.0 \pm 0.1$\\
$ [-45, -40] $& $(-5.4, -12.8)$ & $8.0 \pm 0.2$\\
$ [-40, -35] $& $(-5.8, -12.6)$ & $8.6 \pm 0.2$\\
$ [-35, -30] $& $(-6.3, -11.7)$ & $8.5 \pm 0.2$\\
$ [-30, -25] $& $(-6.9, -11.0)$ & $8.0 \pm 0.2$\\
$ [-25, -20] $& $(-7.0, -10.2)$ & $8.5\pm 0.1$\\
$ [-20, -15] $& $(-7.5, -8.7)$ & $9.2 \pm 0.1$\\
$ [-15, -10] $& $(-7.9, -7.6)$ & $9.2 \pm 0.1$\\
$ [-10,  -5] $& $(-8.5, -6.0)$& $9.7 \pm 0.4$\\
$ [ -5,    0] $& $(-8.0, -4.4)$ & $10.4 \pm 0.2$\\
$ [  0,    5] $& $(-7.6, -3.3)$&  \\\hline

\end{tabular}

\end{table}

\begin{table}\centering
\caption{Sky positions}
	\label{sky_pos}

	\begin{tabular}{cc}
		\hline
		$\phi_1$  &  $ \phi_2$ \\
		$[$deg] & [deg ] \\
		-80& $-1.87 \pm 0.50 $\\
     -70& $ -1.29 \pm 0.40 $\\
     -60& $ -0.66 \pm 0.25 $\\
     -55&  $-0.43 \pm 0.25 $\\
     -50&  $-0.10 \pm 0.25 $\\
     -45& $-0.03 \pm 0.25 $\\
     -40&  $ 0.13 \pm 0.25 $\\
     -35&  $ 0.10 \pm 0.20 $\\
     -30&  $ 0.13 \pm 0.17 $\\
     -25&  $0.06 \pm 0.15 $\\
     -15&  $-0.11 \pm 0.10 $\\
     -10&  $-0.23 \pm 0.10 $\\
     -5&  $-0.71 \pm 0.12 $\\
      0&   $-1.10 \pm 0.15 $\\
      5&   $-1.62 \pm 0.15 $\\\hline

\end{tabular}

\end{table}


\begin{table}\centering
\caption{SDSS spectra}
\label{spec_sdss}

\begin{tabular}{cccccccccccc}  \hline	     
Spec ID & $\alpha$ & $\delta$ & $\phi_1$ & $\phi_2$&  $\mu_{\alpha}cos\delta$& $\mu_{\delta} $ & $g_0$ & $r_0$ & [Fe/H] & RV \\
        & [deg] & [deg] & [deg] &[deg]&  [mas/yr]& [mas/yr] &[mag] & [mag] & [dex] & [km/s]\\\hline
1154-53083-0145 & 126.577155 & -0.439391 & -82.039695 & -1.935876 & -1.23 & -9.80 & 19.16 & 18.96 & -2.33 $\pm$ 0.28 & 246.3 $\pm$ 20.5\\
1154-53083-0155 & 126.645409 & -0.340475 & -81.919882 & -1.946076 & -2.66 & -10.22 & 18.16 & 17.88 & -2.14 $\pm$ 0.23 & 259.7 $\pm$ 14.4\\
1154-53083-0633 & 127.278928 & -0.065362 & -81.366291 & -2.359597 & -2.35 & -10.32 & 19.51 & 19.26 & -2.41 $\pm$ 0.31 & 263.9 $\pm$ 21.6\\
1154-53083-0606 & 127.252030 & 1.086253 & -80.379438 & -1.764341 & -2.18 & -10.54 & 19.81 & 19.54 & -2.26 $\pm$ 0.36 & 248.1 $\pm$ 25.4\\
3293-54921-0483 & 130.382898 & 6.422579 & -74.193561 & -1.804304 & -0.88 & -11.44 & 18.94 & 18.57 & -1.94 $\pm$ 0.31 & 176.4 $\pm$ 18.2\\
5285-55946-0230 & 130.258665 & 7.781532 & -73.080699 & -1.014204 & -2.32 & -11.60 & 18.44 & 18.20 & -2.09 $\pm$ 0.29 & 213.6 $\pm$ 20.7\\
... ... & ...\\\hline
 \end{tabular}	
 Note: Only a portion of table is shown here for illustration. The whole table
contains information of 136 spectra from 116 individual stars are available in the online electronic version. The spectral parameters are calculated by ULySS with ELODIE interpolator. BHB and BS spectra are not given here.

 \end{table}

\begin{table}\centering \tiny
\caption{LAMOST spectra}
\label{spec_lamost}
\begin{tabular}{cccccccccccc}  \hline	     
Spec ID & $\alpha$ & $\delta$ & $\phi_1$ & $\phi_2$&  $\mu_{\alpha}cos\delta$& $\mu_{\delta} $ & $g_0$ & $r_0$ & [Fe/H] & RV \\
        & [deg] & [deg] & [deg] &[deg]&  [mas/yr]& [mas/yr] &[mag] & [mag] & [dex] & [km/s]\\\hline
20170101HD093318N282204M0202028 & 141.661364 & 26.841665 & -51.177836 & -0.194631 & -4.60 & -13.22 & 16.98 & 16.55 & -1.97 $\pm$ 0.11 & 69.2 $\pm$ 5.7\\
20170226HD092331N251058B0112136 & 141.661364 & 26.841665 & -51.177836 & -0.194631 & -4.60 & -13.22 & 16.98 & 16.55 & -2.15 $\pm$ 0.06 & 55.8 $\pm$ 3.1\\
20161126HD091735N272519M0206144 & 141.568500 & 26.967834 & -51.119222 & -0.055559 & -4.65 & -12.96 & 18.11 & 17.76 & -2.45 $\pm$ 0.09 & 56.0 $\pm$ 6.8\\
20170101HD093318N282204M0203077 & 141.779273 & 28.159696 & -50.025682 & 0.453987 & -4.89 & -12.97 & 17.38 & 17.32 & -2.15 $\pm$ 0.03 & 79.3 $\pm$ 3.6\\
20170101HD093318N282204M0215077 & 142.787822 & 29.461088 & -48.452579 & 0.453902 & -4.38 & -13.29 & 17.47 & 17.06 & -2.16 $\pm$ 0.07 & 39.6 $\pm$ 4.4\\
20131113HD093318N282204B0112045 & 144.806430 & 30.124543 & -46.905954 & -0.602918 & -5.11 & -13.10 & 16.60 & 16.13 & -1.80 $\pm$ 0.11 & 31.2 $\pm$ 5.1\\
20170129HD094135N311640B0205233 & 144.806430 & 30.124543 & -46.905954 & -0.602918 & -5.11 & -13.10 & 16.60 & 16.13 & -2.02 $\pm$ 0.05 & 41.5 $\pm$ 2.5\\
20111227F559230403032 & 144.120941 & 30.947720 & -46.574397 & 0.354304 & -4.85 & -13.35 & 17.98 & 17.75 & -2.17 $\pm$ 0.25 & 9.9 $\pm$ 13.9\\
20111220B559160614045 & 145.582829 & 31.777470 & -45.174505 & -0.181384 & -5.19 & -12.69 & 14.94 & 14.56 & -1.86 $\pm$ 0.08 & 11.7 $\pm$ 4.4\\
20130413HD094135N311640F0104221 & 145.582829 & 31.777470 & -45.174505 & -0.181384 & -5.19 & -12.69 & 14.94 & 14.56 & -1.95 $\pm$ 0.04 & 15.3 $\pm$ 2.2\\
20121120HIP4761705133 & 145.582829 & 31.777470 & -45.174505 & -0.181384 & -5.19 & -12.69 & 14.94 & 14.56 & -2.05 $\pm$ 0.04 & 22.2 $\pm$ 2.2\\
20120121F559480311141 & 145.676545 & 32.363379 & -44.653108 & 0.097439 & -5.17 & -12.88 & 16.07 & 15.57 & -2.29 $\pm$ 0.12 & -8.4 $\pm$ 4.9\\
20130413HD094135N311640F0115133 & 145.676545 & 32.363379 & -44.653108 & 0.097439 & -5.17 & -12.88 & 16.07 & 15.57 & -2.28 $\pm$ 0.10 & 11.6 $\pm$ 5.4\\
20170129HD094135N311640B0209118 & 145.676545 & 32.363379 & -44.653108 & 0.097439 & -5.17 & -12.88 & 16.07 & 15.57 & -2.33 $\pm$ 0.04 & 18.4 $\pm$ 1.8\\
20170317HD093848N300911B0112245 & 145.676545 & 32.363379 & -44.653108 & 0.097439 & -5.17 & -12.88 & 16.07 & 15.57 & -2.28 $\pm$ 0.13 & 20.2 $\pm$ 6.4\\
20111227F559230412143 & 146.108792 & 32.808579 & -44.078864 & 0.064429 & -5.03 & -12.83 & 17.40 & 17.01 & -2.24 $\pm$ 0.26 & -8.7 $\pm$ 19.5\\
20131206HD095000N333605M0105199 & 146.108792 & 32.808579 & -44.078864 & 0.064429 & -5.03 & -12.83 & 17.40 & 17.01 & -2.26 $\pm$ 0.06 & 8.8 $\pm$ 2.9\\
20170403HIP48512gw0109157 & 148.804229 & 35.888640 & -40.279970 & 0.131574 & -5.56 & -12.60 & 14.10 & 13.44 & -2.50 $\pm$ 0.03 & -12.8 $\pm$ 1.9\\
20120122F559490405166 & 152.583159 & 39.754518 & -35.397703 & 0.224503 & -6.11 & -12.40 & 17.37 & 16.90 & -2.00 $\pm$ 0.12 & -35.7 $\pm$ 6.9\\
20130306HD102234N423659M0103249 & 153.787485 & 42.230398 & -32.921557 & 1.133103 & -6.18 & -12.08 & 13.91 & 13.24 & -2.38 $\pm$ 0.02 & -68.8 $\pm$ 1.5\\
20130306HD102234N423659B0103249 & 153.787485 & 42.230398 & -32.921557 & 1.133103 & -6.18 & -12.08 & 13.91 & 13.24 & -2.38 $\pm$ 0.03 & -69.3 $\pm$ 2.2\\
20150423HD102234N423659V0103249 & 153.787485 & 42.230398 & -32.921557 & 1.133103 & -6.18 & -12.08 & 13.91 & 13.24 & -2.35 $\pm$ 0.02 & -64.9 $\pm$ 1.2\\
20151231HD104520N453358B0211053 & 162.052840 & 47.145408 & -25.318253 & 0.253396 & -6.98 & -10.66 & 14.98 & 14.43 & -2.23 $\pm$ 0.05 & -121.6 $\pm$ 2.5\\
20180112HD105331N491352M0208053 & 165.199468 & 48.669257 & -22.733558 & -0.044503 & -7.37 & -9.78 & 17.90 & 17.52 & -2.12 $\pm$ 0.08 & -152.8 $\pm$ 4.7\\
20171211HD105355N474016B0112011 & 166.018999 & 49.153350 & -22.009486 & -0.031736 & -7.26 & -9.99 & 16.48 & 15.92 & -2.00 $\pm$ 0.10 & -146.1 $\pm$ 4.7\\
20120218F559761207198 & 175.556867 & 53.177612 & -14.819942 & -0.412337 & -7.35 & -7.73 & 16.69 & 16.21 & -2.29 $\pm$ 0.14 & -208.1 $\pm$ 10.6\\
20120103F559300310163 & 181.925102 & 55.559525 & -10.416717 & -0.249590 & -7.87 & -6.88 & 16.23 & 15.71 & -2.49 $\pm$ 0.08 & -210.6 $\pm$ 4.9\\
20170418HD121217N554221B0203180 & 181.925102 & 55.559525 & -10.416717 & -0.249590 & -7.87 & -6.88 & 16.23 & 15.71 & -2.27 $\pm$ 0.05 & -214.9 $\pm$ 2.3\\
20180207HD121650N561653B0110163 & 181.925102 & 55.559525 & -10.416717 & -0.249590 & -7.87 & -6.88 & 16.23 & 15.71 & -2.36 $\pm$ 0.04 & -206.1 $\pm$ 2.2\\
20120124F559510613168 & 196.468072 & 57.077875 & -2.381764 & -1.845128 & -6.33 & -2.51 & 17.69 & 17.23 & -2.06 $\pm$ 0.21 & -271.4 $\pm$ 10.3\\
20170423HD132545N565813M0215017 & 200.544391 & 58.011729 & -0.033208 & -1.465803 & -7.73 & -3.93 & 18.00 & 17.52 & -2.29 $\pm$ 0.21 & -265.1 $\pm$ 12.8\\
20130521HD140517N563114B0116134 & 209.542501 & 58.440027 & 4.715407 & -1.725496 & -7.49 & -2.51 & 13.73 & 12.98 & -2.31 $\pm$ 0.03 & -279.0 $\pm$ 1.9\\\hline
 \end{tabular}	
 
  Note: BHB and BS spectra are not given here.

\end{table}

\begin{table}\centering
\caption{Four Overdensities of the GD-1 stream}
\label{gd1_od}

\begin{tabular}{ccccccc}  \hline	     
Name & Range &  Center in $\phi_2^c$ & FWHM & FWHM& Area&Areal Density of Dwarfs \\
           & [deg]   & [deg]   &[deg] & [pc]&[deg$^2$]  & [counts/deg$^2$] \\\hline
  O1    &[-54, -43]& 0.039& 0.68 & 91.9 & 7.51 & 25.7\\
  O2    &[-37, -23]&-0.051& 0.35 & 49.4 & 4.84 & 31.9\\
  O3    &[-17, -11]& 0.147&0.28 & 44.6 &1.68 &50.5\\
  O4    &[-3, 2]  &-0.115& 0.58 &104.7 & 2.91 &29.4\\\hline

 \end{tabular}	
 
\end{table}

\begin{table}\centering
\caption{Spectra of BHB and BS stars in LAMOST DR6}
\label{gd1_blue_spec_lamost}
\begin{tabular}{ccccccc}  \hline
 Spec ID & $\alpha$ & $\delta$ & $\phi_1$ & $\phi_2$  & RV &BHB/BS \\
        & [deg] & [deg] & [deg] &[deg]    & [km/s]& \\\hline
20170226HD092331N251058B0103128 & 140.236539 & 25.533824 & -52.974503 & 0.147497 & 84.4 $\pm$ 0.5& BHB \\
20131113HD093318N282204B0115101 & 143.453181 & 29.120593 & -48.403614 & -0.217124 & 45.9 $\pm$ 0.3& BHB \\
20120121F559480310187 & 143.453195 & 29.120636 & -48.403572 & -0.217109 & 16.6 $\pm$ 0.6& BHB \\
20111221F559170516061 & 143.597817 & 29.802689 & -47.771515 & 0.068473 & 35.4 $\pm$ 0.4& BHB \\
20121120HIP4761709243 & 146.897743 & 33.258213 & -43.324801 & -0.202088 & 2.5 $\pm$ 0.3& BHB \\
20170129HD094135N311640B0212185 & 146.897748 & 33.258227 & -43.324787 & -0.202083 & -0.7 $\pm$ 0.4& BHB \\
20160312HD124231N565955B0210048 & 187.384851 & 56.062763 & -7.434435 & -1.125792 & -226.4 $\pm$ 0.4& BHB \\
20130512HD124231N565955M0110031 & 187.384947 & 56.062809 & -7.434368 & -1.125771 & -202.5 $\pm$ 0.7& BHB \\
20130430HD132545N565813F0115016 & 200.544099 & 58.092096 & -0.016401 & -1.387211 & -264.7 $\pm$ 0.3& BHB \\
20161126HD091735N272519M0213010 & 141.779279 & 28.159708 & -50.025669 & 0.453988 & 47.5 $\pm$ 5.9& BS \\
20170101HD093318N282204M0203077 & 141.779279 & 28.159708 & -50.025669 & 0.453988 & 57.3 $\pm$ 0.9& BS \\
20120124F559510406187 & 159.833867 & 45.728513 & -27.402066 & 0.310294 & -36.0 $\pm$ 1.3& BS \\
20130428HD105331N491352M0101031 & 162.557740 & 47.251092 & -24.996359 & 0.094510 & -144.1 $\pm$ 4.3& BS \\
20150120HD104520N453358M0112057 & 162.860793 & 47.482907 & -24.688096 & 0.123543 & -82.6 $\pm$ 5.0& BS \\
20111219F559150601070 & 162.860859 & 47.482989 & -24.688007 & 0.123572 & -108.4 $\pm$ 1.7& BS \\\hline

\end{tabular}	 
\end{table}

\begin{table}\centering
\caption{Spectra of BHB and BS stars in SDSS DR14}
\label{gd1_blue_spec_sdss}
\begin{tabular}{cccccccc}  \hline	   

Spec ID & $\alpha$ & $\delta$ & $\phi_1$ & $\phi_2$ & [Fe/H] & RV &BHB/BS \\
        & [deg] & [deg] & [deg] &[deg]   & [dex] & [km/s]& \\\hline
2889-54530-0215 & 143.453190 & 29.120642 & -48.403570 & -0.217102 & -2.75 $\pm$ 0.14 & 46.4 $\pm$ 1.7& BHB \\
2889-54530-0225 & 143.597840 & 29.802695 & -47.771499 & 0.068460 & -2.07 $\pm$ 0.02 & 41.0 $\pm$ 1.7& BHB \\
3287-54941-0068 & 128.645250 & 3.846764 & -77.291130 & -1.596769 & -2.08 $\pm$ 0.15 & 236.2 $\pm$ 7.1& BS \\
3258-54884-0411 & 155.918420 & 42.360810 & -31.781306 & 0.037050 & -1.82 $\pm$ 0.21 & -64.7 $\pm$ 6.3& BS \\
963-52643-0234 & 160.206670 & 45.502724 & -27.377587 & -0.033768 & -2.25 $\pm$ 0.07 & -97.3 $\pm$ 3.2& BS \\
1018-52672-0247 & 162.557720 & 47.251092 & -24.996369 & 0.094520 & -1.87 $\pm$ 0.02 & -126.6 $\pm$ 2.6& BS \\
2390-54094-0225 & 162.557740 & 47.251092 & -24.996359 & 0.094510 & -1.78 $\pm$ 0.07 & -127.8 $\pm$ 3.7& BS \\
2390-54094-0256 & 162.860870 & 47.482966 & -24.688018 & 0.123550 & -1.48 $\pm$ 0.03 & -131.6 $\pm$ 3.2& BS \\
1186-52646-0137 & 179.937550 & 54.872774 & -11.741092 & -0.305673 & -1.91 $\pm$ 0.15 & -197.8 $\pm$ 9.7& BS \\
1315-52791-0004 & 180.779680 & 54.017752 & -11.731643 & -1.290929 & -1.95 $\pm$ 0.07 & -183.6 $\pm$ 9.8& BS \\
1017-52706-0156 & 186.460310 & 56.627599 & -7.675466 & -0.402289 & -2.13 $\pm$ 0.11 & -186.7 $\pm$ 9.2& BS \\\hline
\end{tabular}

Note: The spectral parameters [Fe/H] and RV are from the database \textsl{sppParams} of SDSS DR14. 
\end{table}

\begin{table}\centering
\caption{BHB and BS stars of the GD-1 stream}
\label{gd1_blue_photo}

\begin{tabular}{cccccccccc}  \hline	     
$\alpha$ & $\delta$ & $\phi_1$ & $\phi_2$ &  $\mu_{\alpha}cos\delta$ & $\mu_{\delta} $ & $g_0$ & $r_0$ & Spec & BHB/BS\\
$[$deg]& [deg] & [deg] &[deg]&  [mas/yr]& [mas/yr] &[mag] & [mag] &  &\\\hline
140.236596 & 25.533821 & -52.974477 & 0.147452 & -4.132 & -13.037 & 14.890 & 15.020 & L & BHB\\
143.453198 & 29.120633 & -48.403573 & -0.217112 & -4.579 & -13.167 & 15.002 & 15.234 & L, S & BHB\\
143.597827 & 29.802693 & -47.771507 & 0.068468 & -4.809 & -13.145 & 14.859 & 14.964 &  L, S& BHB\\
146.897744 & 33.258213 & -43.324801 & -0.202087 & -5.180 & -12.972 & 15.166 & 15.459 &  L& BHB\\
178.682379 & 54.173819 & -12.721899 & -0.544864 & -7.464 & -7.264 & 15.300 & 15.408 &  & BHB\\
187.384947 & 56.062809 & -7.434368 & -1.125771 & -8.011 & -6.015 & 15.519 & 15.741 & L & BHB\\
200.544099 & 58.092096 & -0.016401 & -1.387211 & -7.862 & -3.893 & 15.582 & 15.704 & L & BHB\\
128.645239 & 3.846764 & -77.291135 & -1.596759 & -2.325 & -10.860 & 17.710 & 17.711 & S & BS\\
130.054644 & 4.380349 & -76.125169 & -2.547143 & -0.720 & -9.833 & 16.294 & 16.336 &  & BS\\
141.779273 & 28.159696 & -50.025682 & 0.453987 & -4.887 & -12.968 & 17.384 & 17.319 &S  & BS\\
147.022236 & 34.374419 & -42.366788 & 0.380009 & -5.192 & -13.180 & 17.377 & 17.560 &  & BS\\
150.524843 & 37.541016 & -38.128641 & 0.067106 & -5.820 & -12.536 & 18.349 & 18.294 &  & BS\\
155.918415 & 42.360808 & -31.781310 & 0.037052 & -6.088 & -11.742 & 18.090 & 18.107 & S & BS\\
158.774596 & 46.026867 & -27.705490 & 1.045680 & -6.544 & -11.697 & 18.286 & 18.300 &  & BS\\
159.833866 & 45.728510 & -27.402069 & 0.310293 & -5.701 & -10.161 & 16.840 & 16.869 & L & BS\\
160.206653 & 45.502722 & -27.377597 & -0.033761 & -6.978 & -10.723 & 17.487 & 17.473 &S  & BS\\
162.557771 & 47.251098 & -24.996340 & 0.094501 & -7.011 & -10.948 & 16.959 & 16.919 & L, S & BS\\
162.860856 & 47.482985 & -24.688012 & 0.123571 & -6.854 & -10.264 & 17.170 & 17.176 & L, S & BS\\
171.965059 & 52.272828 & -17.123996 & 0.078116 & -7.347 & -8.635 & 18.718 & 18.645 &  & BS\\
179.937548 & 54.872781 & -11.741089 & -0.305666 & -7.842 & -7.430 & 18.229 & 18.184 & S & BS\\
180.779639 & 54.017737 & -11.731671 & -1.290930 & -8.547 & -7.515 & 18.455 & 18.416 &S  & BS\\
186.460307 & 56.627602 & -7.675466 & -0.402284 & -8.353 & -5.972 & 17.630 & 17.738 & S & BS\\
197.039350 & 57.925555 & -1.860090 & -1.109740 & -8.144 & -4.599 & 18.368 & 18.312 &  & BS\\
197.840722 & 57.840959 & -1.469860 & -1.300677 & -6.826 & -2.321 & 17.706 & 17.774 &  & BS\\
199.283127 & 58.341687 & -0.609431 & -0.997524 & -7.900 & -3.862 & 17.266 & 17.427 &  & BS\\
205.070442 & 58.460079 & 2.391450 & -1.447635 & -7.768 & -2.961 & 18.061 & 18.167 &  & BS\\
205.717415 & 59.492334 & 2.865133 & -0.471724 & -7.466 & -2.065 & 16.682 & 16.782 &  & BS\\
205.887381 & 59.186247 & 2.910395 & -0.786611 & -5.876 & -4.208 & 17.250 & 17.288 &  & BS\\\hline

 \end{tabular}	
 
  Note: "L" indicates that the object has one or more spectra in LAMOST DR6 in Table~\ref{gd1_blue_spec_lamost}; "S" indicates that the object has one or more spectra in  SDSS DR14 in Table~\ref{gd1_blue_spec_sdss}.

\end{table}

\acknowledgments

This research is supported by the National Natural Science Foundation of China (NFSC; 
Grant No. 11673036).  Yue Wu acknowledges support from the NSFC, Grant NO. 11403056.
\par
The authors thank the expert anonymous referee, who provided generous detailed feedback that substantially improved the paper. The authors thank Professor Yu-Qin Chen for useful discussions. The authors thank Professor Xiang-Xiang Xue for recommending the \textit{galpy} to us to fit the GD-1 trace.
\par
Guoshoujing Telescope (the Large Sky Area Multi-Object Fiber Spectroscopic Telescope LAMOST) is a National Major Scientific Project built by the Chinese Academy of Sciences. Funding for the project has been provided by the National Development and Reform Commission. LAMOST is operated and managed by the National Astronomical Observatories, Chinese Academy of Sciences.
%

\software{galpy \citep{bov15};  TopCat \citep{tay05}}

\end{document}